\documentclass[aps,prc,twocolumn,superscriptaddress,showpacs,showkeys,floatfix,nofootinbib,altaffilletter]{revtex4}
\usepackage{xspace}
\usepackage{hyperref}
\usepackage{longtable}

\usepackage[latin2]{inputenc}
\usepackage{indentfirst}
\usepackage{enumerate}

\usepackage{amsmath}
\usepackage{amssymb}
\usepackage[english]{babel}
\usepackage{url}

\newif\ifpdf

\ifx\pdfoutput\undefined

\pdffalse

\else

\pdfoutput=1

\pdftrue

\fi

\ifpdf

\usepackage[pdftex]{graphicx}

\else

\usepackage{graphicx}

\fi

\usepackage{stmaryrd}

\begin{document}
\DeclareGraphicsExtensions{.pdf,.png,.eps,.jpg,.ps}
\cleardoublepage

%\linenumbers
\pagenumbering{roman}

\title{\large \bf Measurements of
                  Production Properties of $K^{0}_{S}$ mesons and $\Lambda$ hyperons \\
                  in Proton--Carbon Interactions at 31~GeV/c}

\begin{abstract}

Spectra of $K^{0}_{S}$ mesons and $\Lambda$ hyperons were 
measured in p+C interactions at 31 GeV/c with 
the large acceptance NA61/SHINE spectrometer at the CERN SPS. 
The data were collected with an isotropic graphite 
target with a thickness of 4\% of a nuclear interaction length. 
Interaction cross sections, charged pion spectra, 
and charged kaon spectra were previously measured using the same data set. 
Results on $K^{0}_{S}$ and $\Lambda$ production in p+C interactions serve as 
reference for the understanding of the enhancement of 
strangeness production in nucleus-nucleus collisions. 
Moreover, they provide important input for the improvement of 
neutrino flux predictions for the T2K long baseline neutrino 
oscillation experiment in Japan.
   
Inclusive production cross sections for $K^{0}_{S}$ and $\Lambda$ 
are presented as a function of laboratory momentum in intervals
of the laboratory polar angle covering the range from 0 up to 240~mrad.
The results are compared with predictions of several hadron production models.
The $K^{0}_{S}$ mean multiplicity in production processes $\langle n_{K^{0}_{S}} \rangle$ and the inclusive cross section for $K^{0}_{S}$ production $\sigma_\mathrm{K^{0}_{S}}$ were measured and amount to 0.127 $\pm$ 0.005 (stat) $\pm$ 0.022 (sys) and 29.0 $\pm$ 1.6 (stat) $\pm$ 5.0~(sys)~mb, respectively.

\end{abstract}

\clearpage

 % Authors in alphabetical order.
\noindent

\author{N.~Abgrall}\affiliation{University of Geneva, Geneva, Switzerland}
\author{A.~Aduszkiewicz}\affiliation{Faculty of Physics, University of Warsaw, Warsaw, Poland}
\author{Y.~Ali}\affiliation{Jagiellonian University, Cracow, Poland}
\author{T.~Anticic}\affiliation{Rudjer Boskovic Institute, Zagreb, Croatia}
\author{N.~Antoniou}\affiliation{University of Athens, Athens, Greece}
\author{J.~Argyriades}\affiliation{University of Geneva, Geneva, Switzerland}
\author{B.~Baatar}\affiliation{Joint Institute for Nuclear Research, Dubna, Russia}
\author{A.~Blondel}\affiliation{University of Geneva, Geneva, Switzerland}
\author{J.~Blumer}\affiliation{Karlsruhe Institute of Technology, Karlsruhe, Germany}
\author{M.~Bogomilov}\affiliation{Faculty of Physics, University of Sofia, Sofia, Bulgaria}
% ...

 \collaboration{\bf The NA61/SHINE Collaboration}
 \noaffiliation

 \date{\today}

 \maketitle
 \clearpage

  \pagenumbering{arabic}

%____________________________________________________________________________ 

 \section{Introduction}

Experimental data on strange particle production in proton-proton and 
proton-carbon interactions 
in the region below few tens of GeV/c incident momentum are 
scarce~\cite{d1,d2,d3,d4}. 
There are at least two reasons why the knowledge of yields of strange mesons and baryons are of considerable interest at the beam momentum of 31~GeV/c. One is the need to include the tuned production of these particles in the precise neutrino flux calculation for accelerator neutrino experiments as an additional source of neutrinos and secondary pions. Production of neutral kaons is important for accurate calculation 
of the $\nu_{e}$ and $\bar \nu_{e}$ flux 
from $K^{0}_{L} \rightarrow \pi e \nu_{e}$ decays. 
The other is the understanding of the production of strangeness 
in nucleus-nucleus interactions 
in this energy region and its proposed interpretation 
as a signal of the onset of deconfinement. Hadron-nucleus interactions constitute an intermediate step between proton-proton and nucleus-nucleus interactions. Therefore their study permits to understand better the influence of nuclear matter on strangeness production and can help to understand the role of the nucleus and non-exotic mechanisms of strangeness production. In addition to the primary goal of providing reference data for the T2K neutrino experiment in Japan, precise results on particle production in p+C interactions furnish important input to improve hadronic event generators which are required for the interpretation of air showers initiated by ultra high energy cosmic particles. The NA61/SHINE collaboration already published results on charged pion~\cite{NA61_pion} and $K^{+}$ production~\cite{NA61_Kplus}. In this paper we present results on $K^{0}_{S}$ and $\Lambda$ yields. 
These were obtained from the analysis of the same sample of p+C collisions at 31~GeV/c 
beam momentum collected with the NA61/SHINE large acceptance spectrometer 
at the CERN SPS in 2007.
The statistics of this data sample is insufficient to obtain results
on $\overline{\Lambda}$ yields, which are about 100 times smaller
than $\Lambda$ yields~\cite{Abt:2009aa}.

 	The paper is organized as follows. Section~\ref{Sec:set-up} provides 
information about the NA61/SHINE experimental apparatus. 
The main components of the detector and trigger system are presented. 
In addition, the target used is described 
and numbers  of registered proton interactions are provided. 
The analysis technique is discussed in Section~\ref{Sec:analysis}. 
In this Section, the procedure of the extraction of 
the $K^{0}_{S}$ mesons and $\Lambda$ hyperons is described in detail. The calculation of corrections and the normalization procedure are discussed. Subsection~\ref{Sec:binning} presents event and track selection criteria as well as the adopted phase space binning scheme. Information about extraction of the $K^{0}_{S}$ mesons and $\Lambda$ hyperons from the data and calculation of the correction factors used to correct raw yields for detector and other effects are described in Subsections~\ref{Sec:raw_data} and~\ref{Sec:cor_factors}, respectively. The main sources of the systematic uncertainty are discussed in Subsection~\ref{Sec:Sys}. The final results are shown in Section~\ref{Sec:results}. They are compared to predictions of different hadron production models in Section~\ref{Sec:MC}. Finally, Section~\ref{Sec:Summary} summarizes the results of the paper.

%____________________________________________________________________________ 

\section{The NA61/SHINE experimental set-up}
\label{Sec:set-up}

\begin{figure*}[!ht]
\begin{center}
\includegraphics[width=0.7\linewidth]{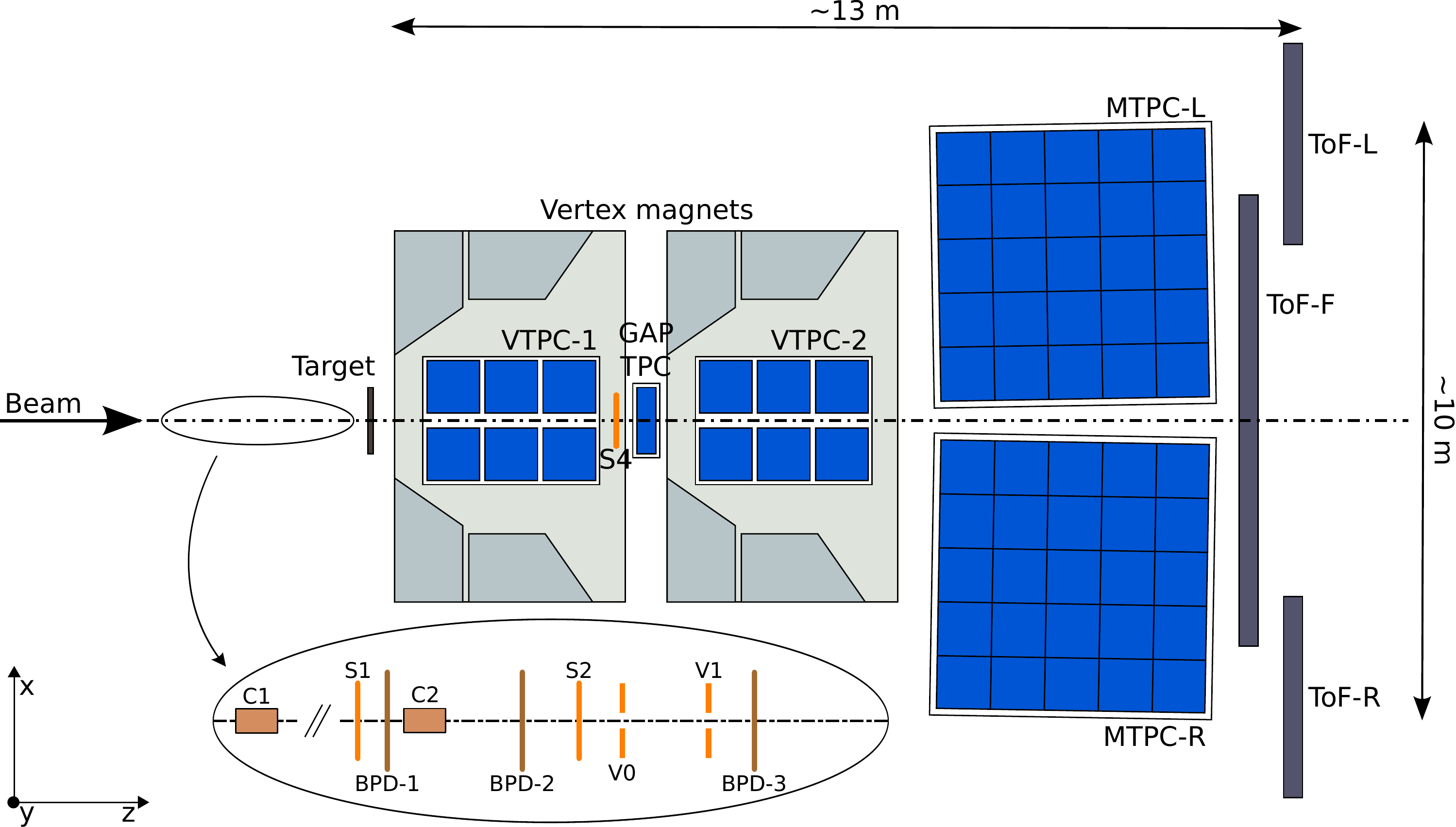}
\end{center}
  \caption{The NA61/SHINE experimental apparatus, see text for details.}
\label{fig:detector}
\end{figure*}

The NA61/SHINE experiment~\cite{proposala} is 
situated in the North Area H2 beam-line of the CERN SPS. 
It is the successor of the NA49 experiment~\cite{na49nim}. 
The NA61/SHINE detector is a large acceptance hadron spectrometer, 
which consists of a set of large volume Time Projection Chambers (TPCs) 
and Time-of-Flight (ToF) scintillator walls. The schematic layout of the detector is shown in Fig.~\ref{fig:detector} together with the overall dimensions.
Two of the TPCs (VTPC-1 and VTPC-2) are placed in the magnetic field produced by two
super-conducting dipole magnets with maximum 
combined bending power of 9~Tm. 
During the 2007 data taking period the magnetic field was 
set to a bending power of 1.14~Tm to optimize the detector acceptance for 
the measurements needed for the T2K experiment.
The TPCs were filled with Ar:CO$_2$ gas mixtures in proportions 90:10 for the VTPCs and 95:5 for the MTPCs. A set of scintillation and Cherenkov counters as well as beam position detectors (BPDs) upstream of the main detection system provided timing reference, identification and position measurements of the incoming beam particles. The 31~GeV/c secondary hadron beam was produced from 400~GeV protons extracted from the SPS in slow extraction mode. The secondary beam was transported along the H2 beam-line towards the experiment. Collimators in the beam line were adjusted to get an average beam particle rate of 15~kHz. Protons in the beam were identified by two Cherenkov counters, a CEDAR and a threshold counter, labeled C1 and C2, respectively. Two scintillation counters, S1 and S2, together with two veto counters, 
V0 and V1, were used to select beam particles. 
The S1 counter also provided the timing (start time for all counters). Beam protons were selected by the coincidence
$\textrm{S1}\cdot\textrm{S2}\cdot\overline{\textrm{V0}}\cdot
\overline{\textrm{V1}}\cdot\textrm{C1} \cdot\overline{\textrm{C2}}$.
The trajectory of individual beam particles is measured in the BPDs along the beam line. These counters are small (3$\times$3~cm$^2$)
proportional chambers with cathode strip readout (BPD-1/2/3 in Fig.~\ref{fig:detector}). A special run was taken to measure the beam momentum by bending the incoming beam particles into the TPCs with the full magnetic field. From this measurement 
the mean momentum value of  30.75~GeV/c was obtained. Interactions in the target were selected by an anti-coincidence of the incoming beam protons with a small, 2~cm diameter, scintillation counter (S4) placed on the beam trajectory between the two vertex magnets (see Fig.~\ref{fig:detector}). This minimum bias trigger was based on 
the disappearance of the incident proton. 
The results presented here were obtained from 667$\times10^{3}$ proton 
interactions recorded with an isotropic graphite target of 
dimensions 2.5(W)$\times$2.5(H)$\times$2(L)~cm$^3$ 
and with a density of $\rho = 1.84$~g/cm$^3$. 
For normalization purposes 46$\times10^{3}$ proton interactions with 
the carbon target removed were also recorded.
The carbon target 
was installed 80~cm in front of VTPC-1. 

%____________________________________________________________________________ 

\section{Analysis Technique}
\label{Sec:analysis}

The most frequent decays of $K^{0}_{S}$ mesons and $\Lambda$ hyperons lead to the production of two oppositely charged particles. 
The measurement of particle tracks in the magnetic field allows to determine 
their charges and momenta. This section presents the method of  $K^{0}_{S}$ and $\Lambda$ analysis using invariant mass distributions. When the $K^{0}_{S}$ hypothesis is studied, positively (negatively) charged tracks are assumed to be $\pi^{+}$  ($\pi^{-}$) mesons. For the $\Lambda$ hypothesis the positively (negatively) charged particles are assumed to be protons ($\pi^{-}$ mesons). The analysis was made in specific invariant mass windows. 
The selected window should cover the invariant mass peak of $K^{0}_{S}$ ($\Lambda$) but also include regions below and above the peak for the background estimate. 
Fits of the background function depend somewhat on the selected side regions. 
This effect was checked and added to the systematic uncertainties (see Sec.~\ref{Sec:Sys}). 
The number of $K^{0}_{S}$ and $\Lambda$ was determined from fitting a sum of Lorentzian function and polynomial function for the signal and the background, respectively. 
The Lorentzian function is described by 
\begin{equation}
f(x)=A \frac{\frac{1}{2}F}{(x-x_{0})^{2}+(\frac{1}{2}F)^{2}}  ~,
\label{Eq:Lorentz_function} 
\end{equation}
where parameter $A$ controls the height of the peak, $F$ is the full width at half maximum (FWHM), and $x_{0}$ is the mean value (in this case, the fitted $K^{0}_{S}$ or $\Lambda$ mass). The low statistics data forced to constrain the width of the signal function according to the Monte Carlo predictions. 
Namely, the width from Monte Carlo was set as initial value. 
Then, this parameter was allowed to vary 
(for $K^{0}_{S}$ up to $\pm$ 5$\cdot$$\Delta F_\mathrm{MC}$, where $\Delta F_\mathrm{MC}$ is an error of 
the fitted Monte Carlo width in a given \{$p$, $\theta$\} bin; 
for $\Lambda$ up to $\pm$7$\cdot$$\Delta F_\mathrm{MC}$). 
The influence of this assumption on the final result was checked and added to the systematic uncertainties  (see Sec.~\ref{Sec:Sys}).   
In the standard approach a 4$^{th}$ order polynomial was used as the background function. The sensitivity to different shapes of the background function was studied and is included in the systematic uncertainties (see Sec.~\ref{Sec:Sys}). The raw number of $K^{0}_{S}$ and $\Lambda$ was calculated as the integral of the fitted signal function~Eq.~\ref{Eq:Lorentz_function}. Corrections were applied to the raw numbers of $K^{0}_{S}$ and $\Lambda$ in order to obtain their yields produced in the primary p+C interactions after strong and electromagnetic decays. Correction factors were derived from a Monte Carlo procedure in which events were generated from the hadron production model VENUS~4.12~\cite{Venus412} and then sent through a full simulation of the detector. The procedure takes into account the trigger bias, the vertex fit requirement and cuts, the branching ratio for the studied type of decay, the geometrical acceptance, the reconstruction efficiency and feed-down from interactions with the target material. The feed-down correction also takes into account $\Lambda$ hyperons coming from secondary vertices. The inverse multiplicative Monte Carlo correction is calculated using the following formula
\begin{equation}
E(p, \theta)= \left ( \frac{\mathrm n_\mathrm{x}}{\mathrm N} \right )^{\mathrm {MC}}_{\mathrm{acc}}  \big / \left (  \frac{\mathrm  n_\mathrm{x}}{\mathrm N} \right )^{\mathrm{MC}}_{\mathrm{gen}}
\label{Eq:correction} 
\end{equation}
which compares the information on simulated particles at the primary hadron 
generator level (gen) with that on reconstructed identified particles (acc). 
The quantity $\mathrm n_\mathrm{x}$ is the number of the identified particle of type $\mathrm x$ 
($K^{0}_{S}$ or $\Lambda$) in a given bin of phase-space, 
$\mathrm N$ is the number of events. The correction can be split into two parts: the first one connected with correction of numbers of a given particle of type $\mathrm x$ in a given bin ($ \mathrm n_\mathrm{x}^{\mathrm{acc}}/\mathrm n_{\mathrm x}^{\mathrm{gen}}$ which will be denoted $\gamma$) and the second one connected with the correction of number of events ($\mathrm N^{\mathrm{acc}}/\mathrm N^{\mathrm{gen}}$ which will be denoted $\eta$). Therefore, the correction can be rewritten as follows:

\begin{equation}
E(p, \theta)= \left ( \frac{\mathrm n_\mathrm{x}^{\mathrm{acc}}}{\mathrm n_\mathrm{x}^{\mathrm{gen}}} \right )^{\mathrm{MC}}  \big / \left (  \frac{\mathrm N^{\mathrm{acc}}}{\mathrm N^{\mathrm{gen}}} \right )^{\mathrm{MC}} = \frac{\gamma}{\eta}~.
\label{Eq:correction_2} 
\end{equation}
The subtraction of non-target interactions was 
performed using information from events recorded with target removed.
The normalization was obtained far away from the target 
where all reconstructed vertices originate from interactions 
with the detector material (neglecting the beam attenuation in the target). 
The number of particles per event in a given phase space bin, 
corrected for non-target interactions, is calculated as:
\begin{equation}
\frac{\mathrm n}{\mathrm N} = \frac{1}{\mathrm E} \frac{\mathrm n^{\mathrm I} -\mathrm B \mathrm n^{\mathrm R}}{\mathrm N^{\mathrm I}-\mathrm{B N}^{\mathrm R} }~,
\label{Eq:norm_1} 
\end{equation}
where I and R superscripts indicate data with target inserted and removed, respectively.  
The factor B is calculated from the data:
\begin{equation}
\mathrm B = \frac{\mathrm N^{\mathrm I}_\mathrm{ beam}}{\mathrm N^{\mathrm R}_\mathrm{beam}} = \frac{\mathrm N^{\mathrm I}_\mathrm{far~z}}{\mathrm N^{\mathrm R}_\mathrm{far~z}}~,
\label{Eq:norm_2} 
\end{equation}
where $\mathrm N^\mathrm{R/I}_\mathrm{beam}$ is the number of beam particles in data with target removed and inserted, respectively, and $\mathrm N^\mathrm{R/I}_\mathrm{far~z}$ is the corresponding number of events with fitted vertex longitudinal coordinate, z, far away from the target.
The differential spectrum is calculated as
\begin{equation}
\frac{\mathrm {dn}}{\mathrm {dp}}=\frac{\mathrm n}{\mathrm N} \frac{1}{\mathrm \Delta p}~.
\label{Eq:norm_3} 
\end{equation}
Then the differential cross section is calculated as:
\begin{equation}
%\frac{d\sigma}{dp}=\sigma_{prod} \frac{dn}{dp}~,
\frac{\mathrm d\sigma}{\mathrm {dp}}=\sigma_\mathrm{prod} \frac{\mathrm {dn}}{\mathrm {dp}}~,
\label{Eq:norm_4} 
\end{equation}
where $\sigma_\mathrm{prod}$ is equal to 229.3 $\pm$ 1.9 $\pm$ 9.0 mb. The 
uncertainties on $\sigma_\mathrm{prod}$  were not included in the uncertainties of
the final results presented in this paper. 
The measurements of the inelastic and production cross sections 
are presented in Ref.~\cite{Claudia}.

%____________________________________________________________________________ 

\subsection{Event and track selection, data binning}
\label{Sec:binning}

The analysis is based on 667$\times10^{3}$ event triggers with 
the graphite target inserted and 46 $\times10^{3}$ triggers with the target removed. 
Only events with a properly reconstructed beam track were retained. 
First of all, the information from the Beam Position Detectors, 
placed upstream of the target (see Fig.~\ref{fig:detector}), 
was used to ensure a well-defined beam track (see Ref.~\cite{proposala} for details). 
Then, events with reconstructed primary vertex, within the target 
(vertex $z$ position within the range $[-585.0, -575.0]$~cm), were selected in order to reject
interactions that did not take place in the target but in the surrounding detector
material. Any pair of tracks with opposite charges and distance of closest 
approach smaller than 1~cm  was taken as a possible  $V^{0}$ candidate. 
In order to purify the sample and select candidates which correspond 
to $K^{0}_{S}$ or $\Lambda$ with high probability 
additional cuts were applied:

\begin{enumerate}[(i)]

\item{The distance along the beam direction between the primary vertex  and the $V^{0}$ decay point had to be larger than 3 cm. 
This cut was used to reject cases in which primary vertex tracks 
were wrongly reconstructed as $V^{0}$ tracks. The same minimum distance cut 
was used for $K^{0}_{S}$ and $\Lambda$. Although the decay lengths of the two particles are different, the signal to background ratio are very similar using the same cut.}

\item{The $V^{0}$ momentum vector had to point back to the primary vertex within a cut of $d_x$ and $d_y < 3$~cm in the transverse \{$x,y$\} plane for $K^{0}_{S}$ and $\Lambda$ candidates.} 

\item{The analysis was performed 
in the invariant mass windows  $[0.35, 0.7]$~GeV/$c^2$ 
and  $[1.09, 1.16]$~GeV/$c^2$ for $K^{0}_{S}$ and $\Lambda$, respectively.}

\item{A cut was applied on the angle between the momentum of the $V^{0}$ candidate and its decay products 
in the c.m. system. Due to the expected isotropy of the decay in the $V^{0}$ rest frame the 
distribution of the cosine of this angle (cos~$\epsilon$) should be flat. 
The cut was set to -0.95$<$cos~$\epsilon$$<$0.8 and -0.8$<$cos~$\epsilon$$<$0.8 for $K^{0}_{S}$ and $\Lambda$, respectively. 

\item{Cuts in the Armenteros-Podolanski plot were applied. 
The relation between Armenteros transverse momentum 
($\mathrm p^{\mathrm arm}_\mathrm{T}$) and longitudinal momentum asymmetry ($\alpha$) 
can be used to separate $K^{0}_{S}$ mesons from $\Lambda$ 
hyperons with high probability. The separation of $K^{0}_{S}$ and $\Lambda$ 
candidates using the Armenteros-Podolanski plot is shown in Fig.~\ref{fig:K0S_armenteros}.}
For the selection of $K^{0}_{S}$ candidates, 
the regions $\mathrm p^\mathrm{arm}_\mathrm{T}$ and $<$0.05 were excluded. 
In addition the region $\alpha$$>$0.5 and $\mathrm p^\mathrm{arm}_\mathrm{T}$$<$ 0.12 is rejected in order to get rid of background originating from $\Lambda$ particles.
For the $\Lambda$ selection, the region $\mathrm p^\mathrm{arm}_\mathrm{T}$$<$0.03 or $>$0.11 was excluded. 
Furthermore, the region $\alpha$$<$0.45 was rejected.  
\begin{figure}[h!]
\begin{center}
\includegraphics[width=0.70\linewidth]{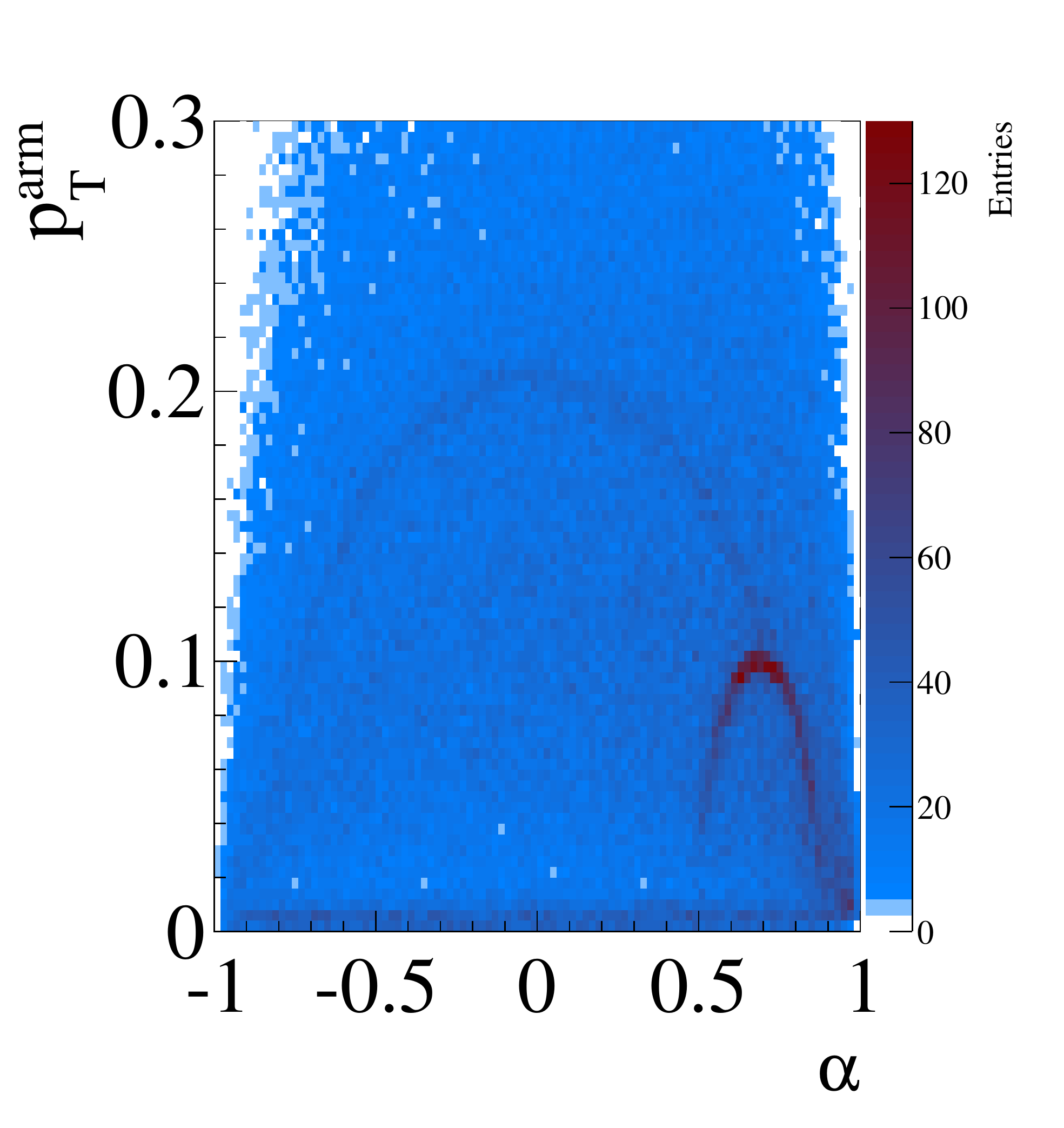}
\includegraphics[width=0.45\linewidth]{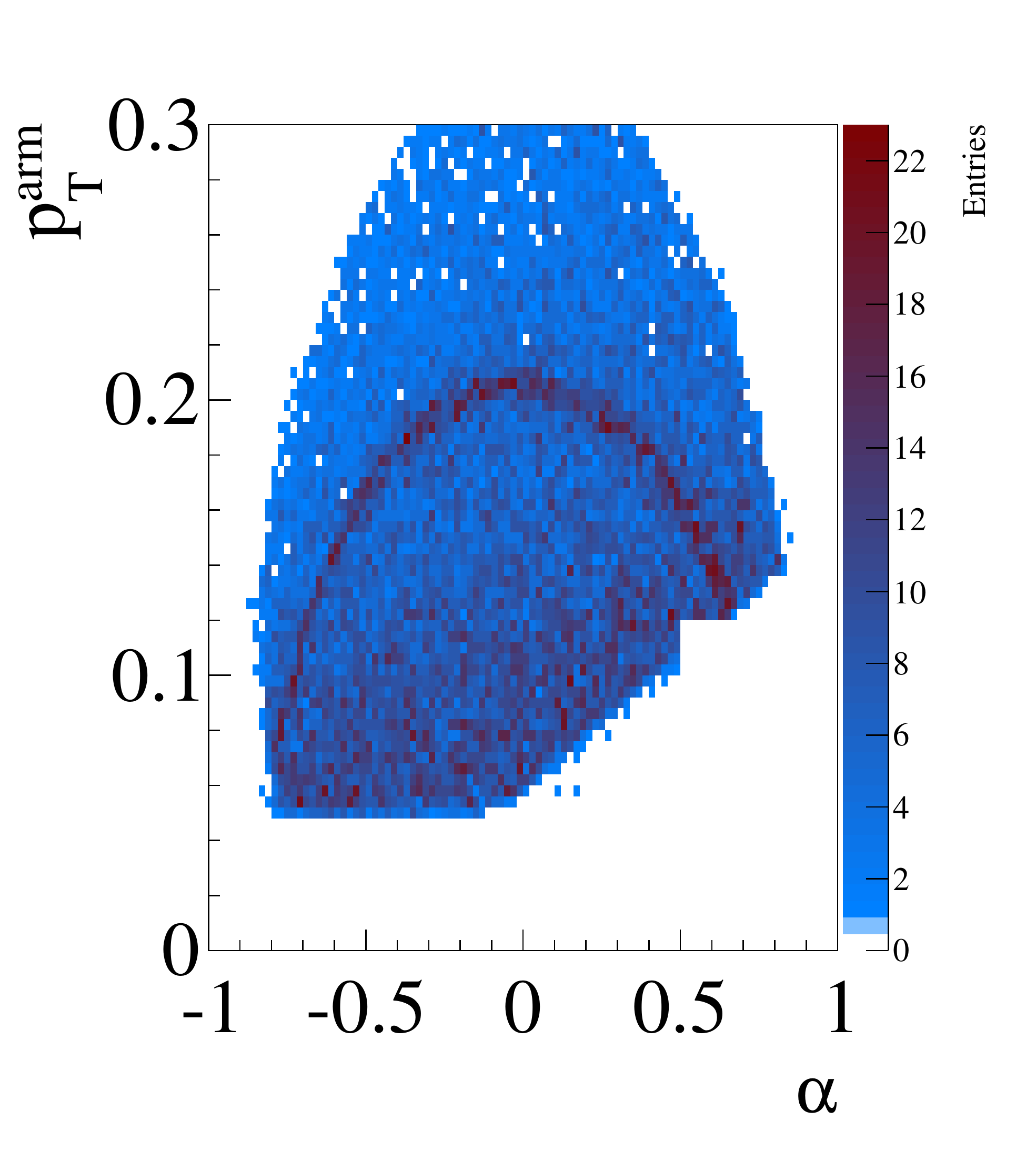}
\includegraphics[width=0.45\linewidth]{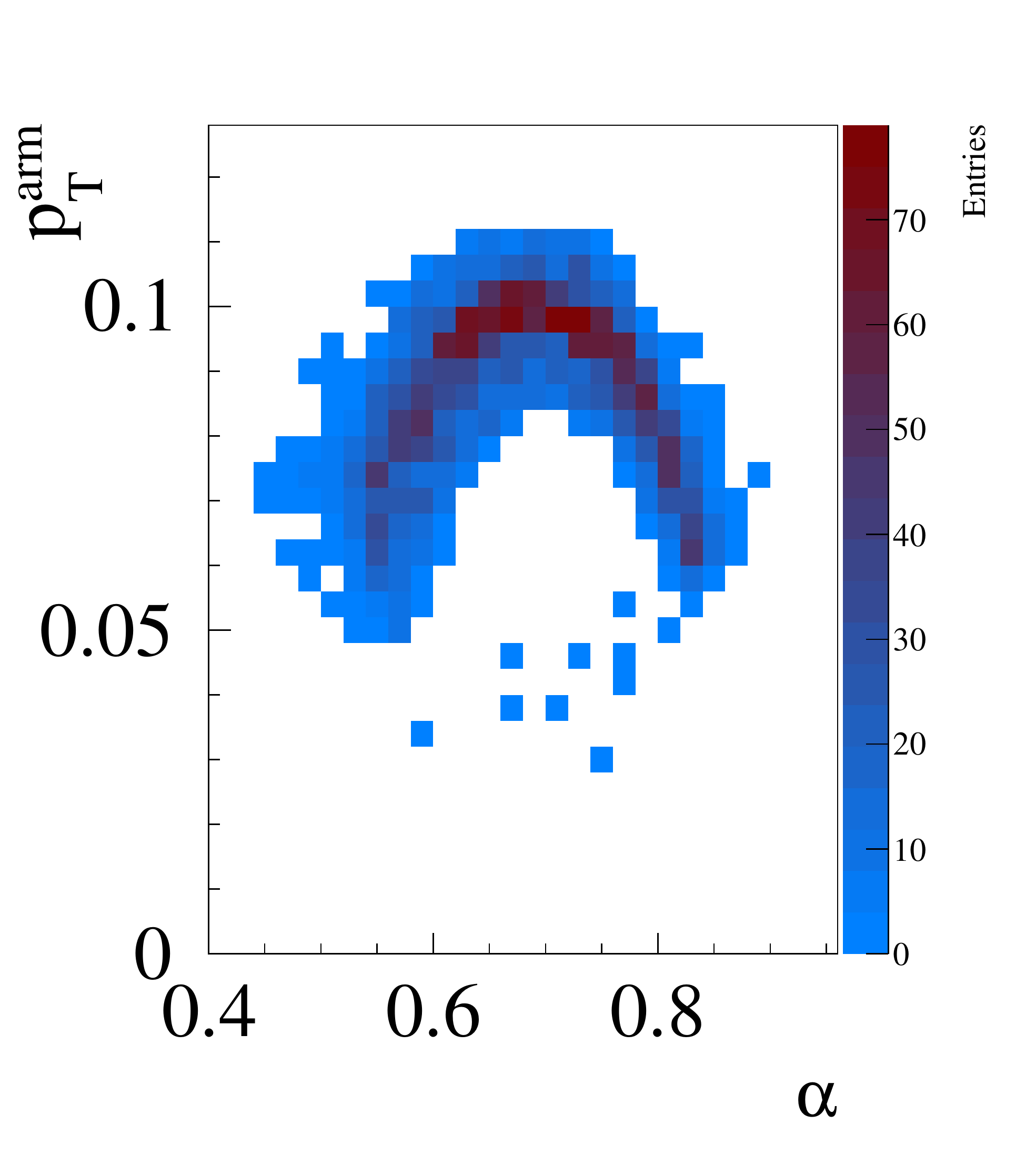}
\end{center}
  \caption{Armenteros-Podolanski plots before (top) and after 
all cuts for $K^{0}_{S}$ (bottom left) and $\Lambda$ candidates (bottom right).}
\label{fig:K0S_armenteros}
\end{figure}
} 
\end{enumerate}

\begin{figure}[!h]
%\begin{figure*}[!h]
\begin{center}
\includegraphics[width=0.9\linewidth]{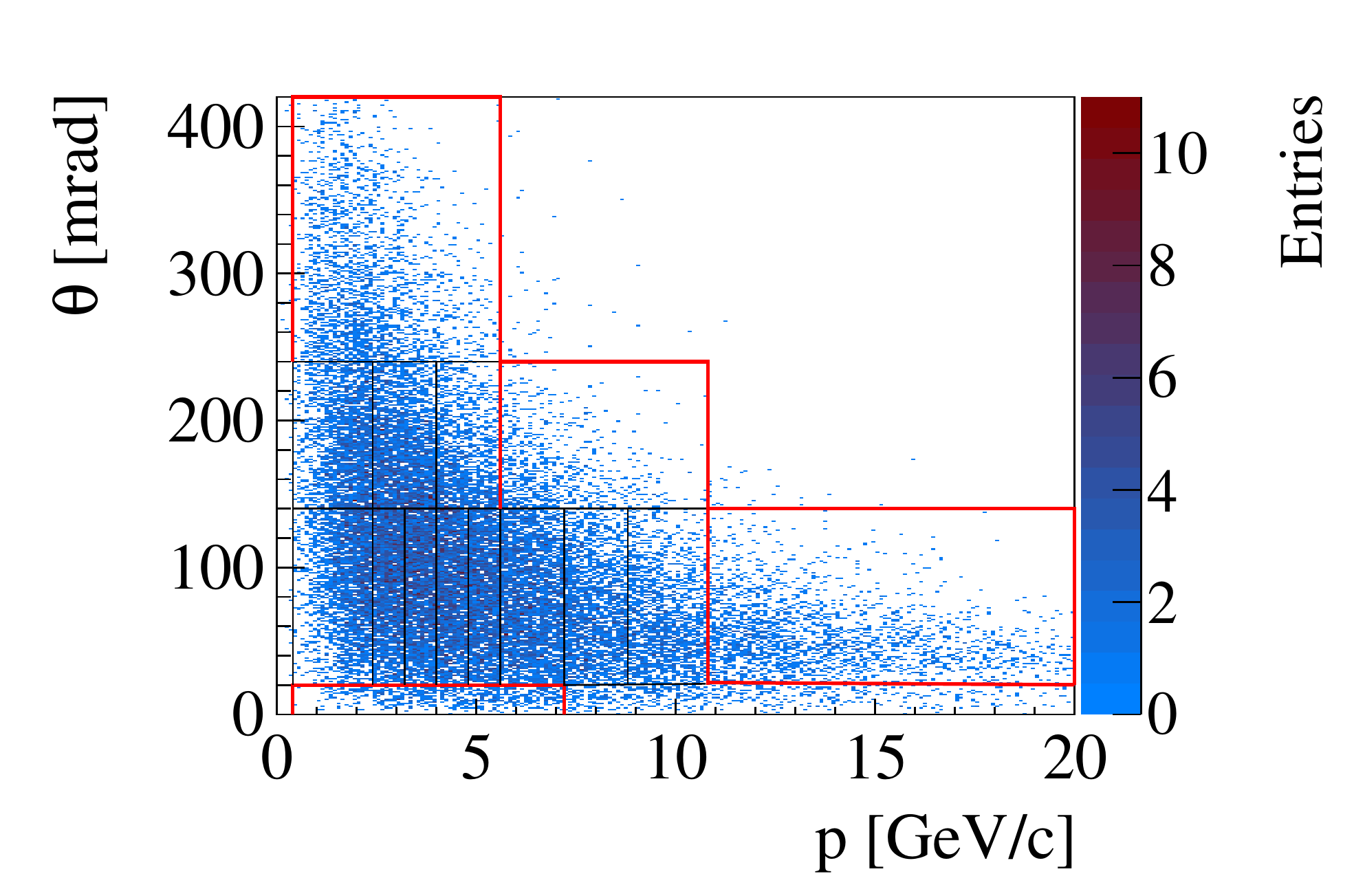}
\includegraphics[width=0.9\linewidth]{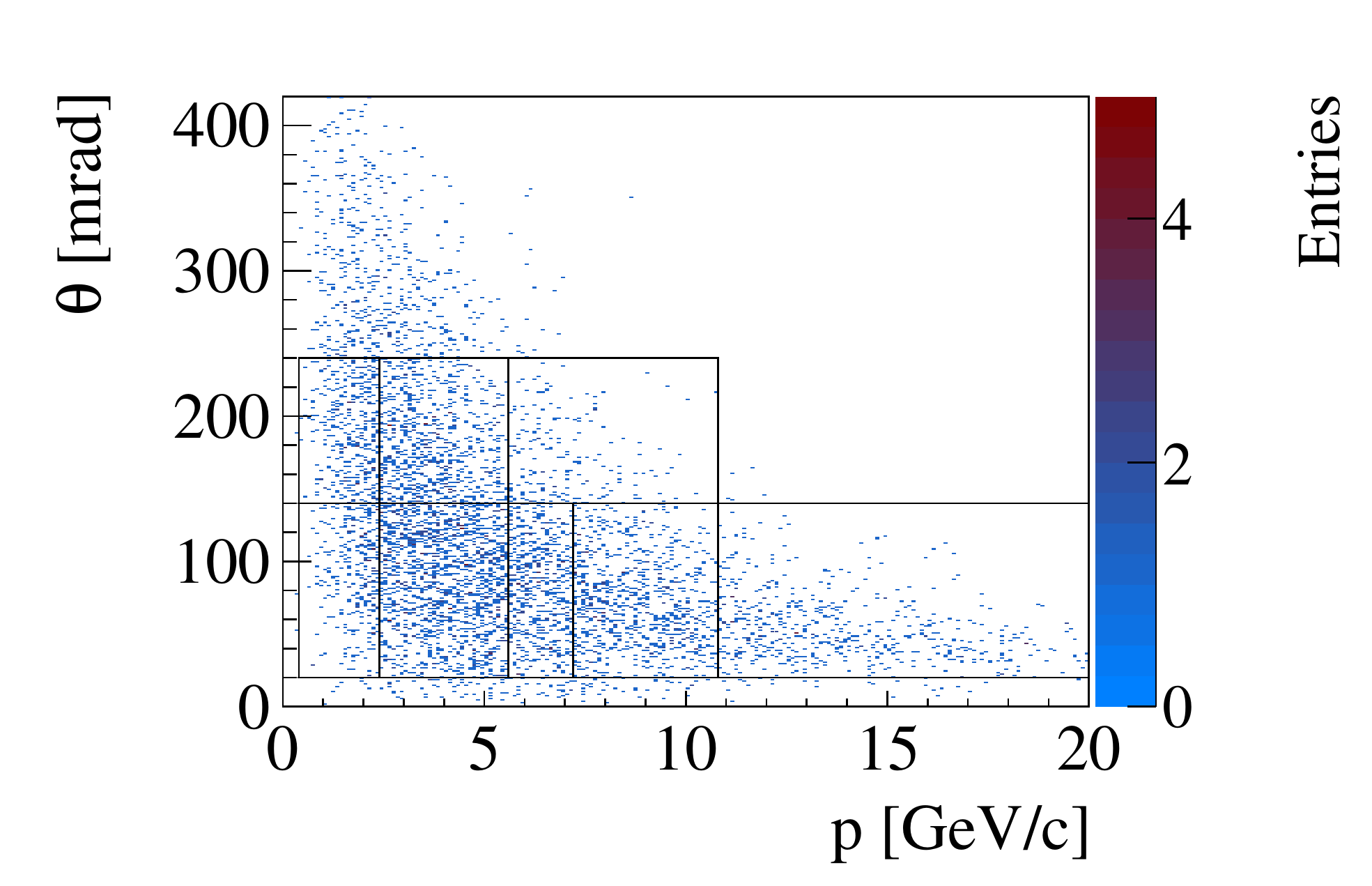}
\end{center}
  \caption{The momentum vs polar angle distributions for $K^{0}_{S}$ (top) and $\Lambda$ (bottom) candidates with superimposed binning.}
\label{fig:K0S_binning}
\end{figure}

The $K^{0}_{S}$ candidate momentum vs polar angle distribution after event, track and $V^{0}$ selection cuts is shown in Fig.~\ref{fig:K0S_binning} with superimposed binning. The bins colored in red were more sensitive to the model dependent corrections 
because for them the uncertainty of the shape of the event generator 
distributions could significantly affect the final results. 
   
%\end{figure*}

%____________________________________________________________________________ 

\subsection{Raw Yields}
\label{Sec:raw_data}

%\begin{figure}[!h]
\begin{figure*}[!h]
\begin{center}
\includegraphics[width=0.95\linewidth]{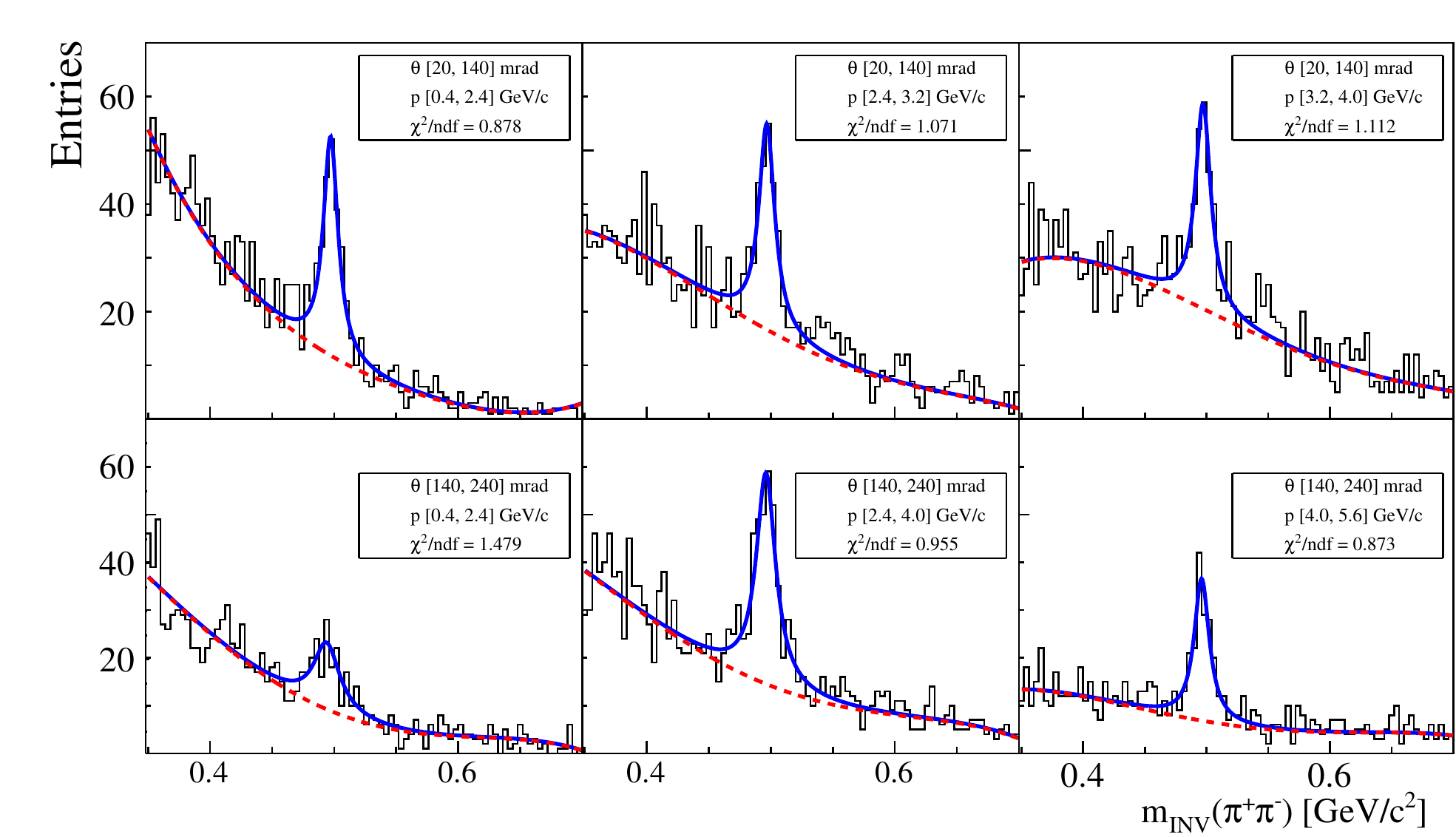}
\end{center}
  \caption{The  invariant mass distributions for $K^{0}_{S}$ candidates 
in selected \{$p$, $\theta$\} bins. 
Data (black) with superimposed background function (red dashed) 
and global fit (blue solid) are shown.}
\label{fig:K0S_minv}
%\end{figure}
\end{figure*}

\begin{figure*}[!h]
\begin{center}
\includegraphics[width=0.99\linewidth]{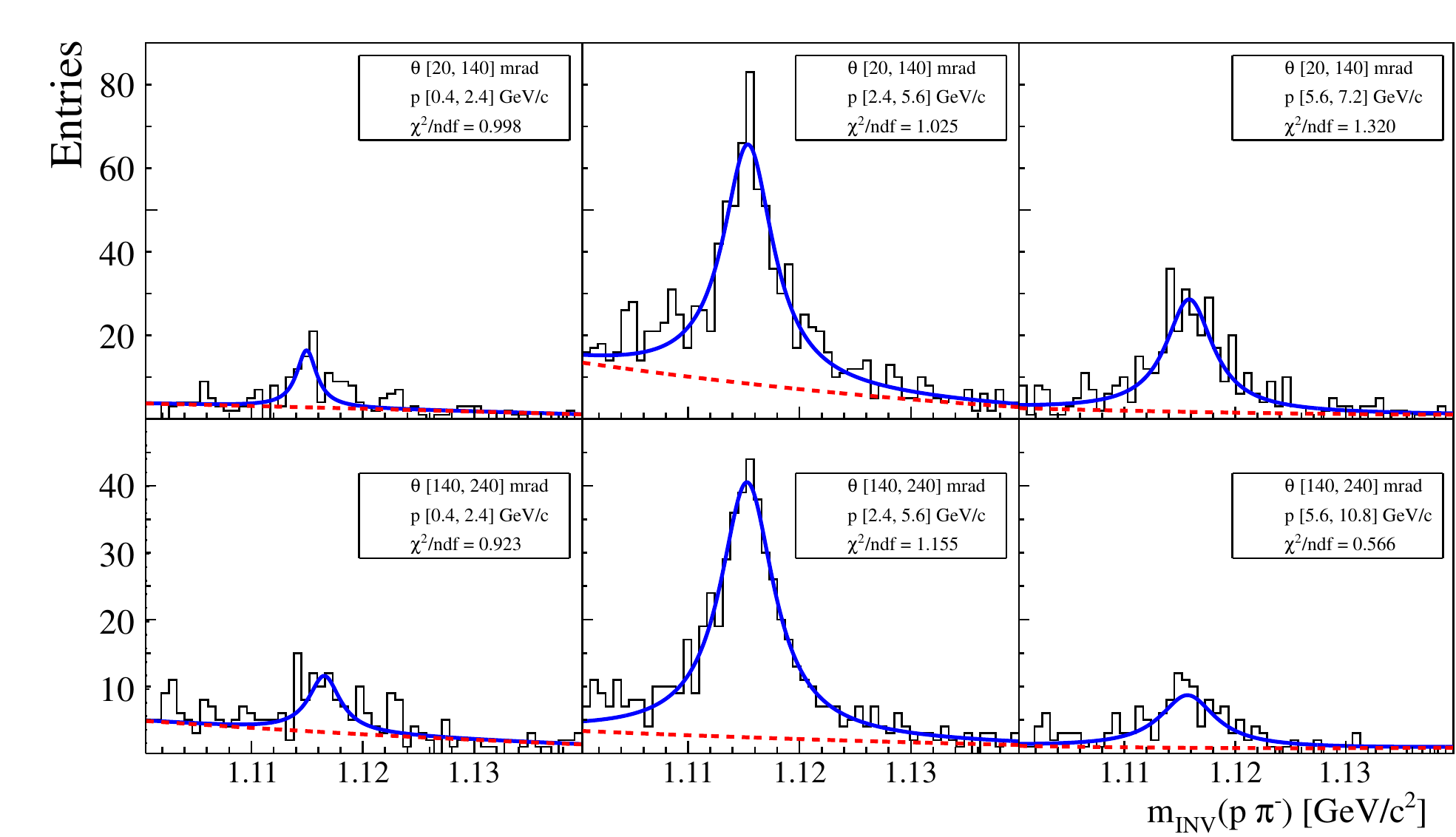}
\end{center}
  \caption{The invariant mass distributions for $\Lambda$ candidates 
in selected \{$p$, $\theta$\} bins. 
Data (black) with superimposed background function 
(red dashed) and global fit (blue solid) are shown.}
\label{fig:Lambda_minv}
\end{figure*}

The ($\pi^{+}$$\pi^{-}$) invariant mass distributions in selected \{$p$, $\theta$\} bins for $K^{0}_{S}$ candidates 
are presented in Fig.~\ref{fig:K0S_minv}.
The invariant mass distributions for $\Lambda$ candidates
in selected \{$p$, $\theta$\} bins are presented in Fig. \ref{fig:Lambda_minv}.
The raw number of $K^{0}_{S}$ and $\Lambda$ extracted in the selected momentum and polar angle intervals is presented in Table \ref{tab:K0S_data}.
\begin {table}[h!]
\caption{
\label{tab:K0S_data}
The raw number of $K^{0}_{S}$ and $\Lambda$
extracted by the fitting procedure in selected momentum and polar angle intervals}
\begin{tabular}{l l|l l|l|l||l l|l|l}
\hline
\multicolumn{6}{c||}{$K^{0}_{S}$} &
\multicolumn{4}{c}{$\Lambda$} \\
$\mathrm \theta_\mathrm{low}$ & $\mathrm \theta_\mathrm{up}$ & $\mathrm p_\mathrm{low}$ & $\mathrm p_\mathrm{up}$ & n & $\mathrm \Delta \mathrm n$ & $\mathrm p_\mathrm{low}$ & $\mathrm p_\mathrm{up}$ & n & $\mathrm \Delta \mathrm n$ \\

\multicolumn{2}{c|}{(mrad)} & \multicolumn{2}{c|}{(GeV/c)} &   &                  &  \multicolumn{2}{c|}{(GeV/c)} &       &   \\

\hline \hline
  0& 20	& 0.4   &7.2    &61    &10  & &  & &\\
\hline
  20&140	&0.4    &2.4    &178    &23 & 0.4 & 2.4 & 53& 8	\\
  &	        &2.4    &3.2    &167    &18 & 2.4& 5.6 &562 &26	\\
 &	        &3.2    &4.0    &165    &18 & & & &	\\
 &	        &4.0    &4.8    &126    &18 & & & &	\\
&	        &4.8    &5.6    &132    &16 & & & &	\\
&	        &5.6    &7.2    &130    &16 & 5.6&7.2 &248 &16	\\
&	        &7.2    &8.8    &80    &15	 & 7.2&10.8 &468 &22\\
&	        &8.8    &10.8    &115    &15 & & & &	\\
&	        &10.8    &20.0    &102    &17 & 10.8&20.0 &362 &19	\\
\hline
  140&240	&0.4    &2.4    &93    &15 & 0.4& 2.4&60 &12	\\
 &	                &2.4    &4.0    &233    &20 & 2.4&5.6 &344 &20	\\
  &	                &4.0    &5.6    &135    &20 & & & &	\\
  &	                &5.6    &10.8    &88    &12 & 5.6& 10.8& 72 &17	\\
\hline 
 240&420	&0.4    &5.6    &135    &26 & & & &	\\

\hline
\end{tabular}
\end {table}

%____________________________________________________________________________ 

\subsection{Corrections}
\label{Sec:cor_factors}

\begin{figure}[!h]
%\begin{figure*}[!h]
\begin{center}
\includegraphics[width=0.8\linewidth]{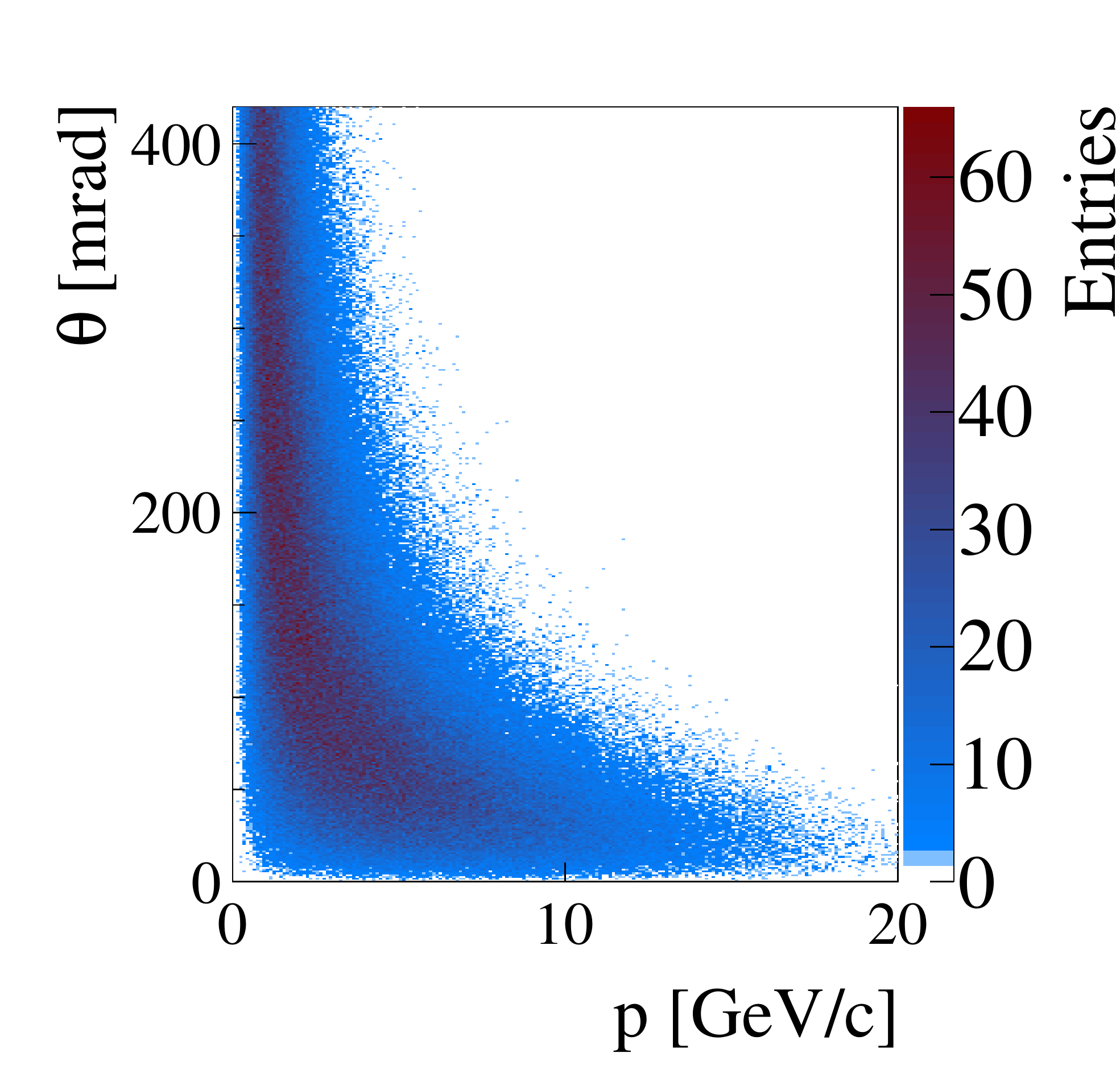}
\includegraphics[width=0.8\linewidth]{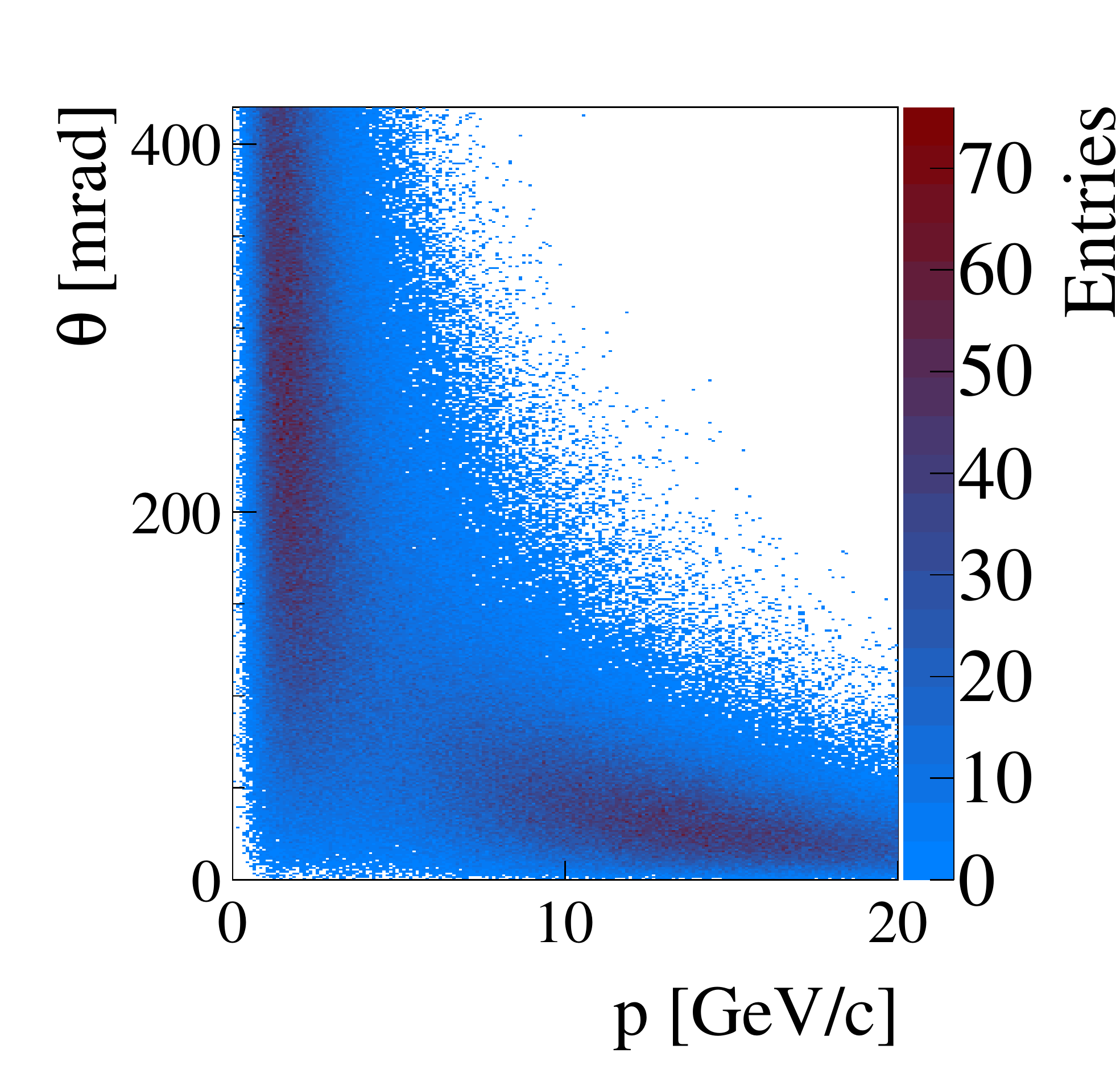}
\end{center}
  \caption{The momentum versus polar angle distribution for 
simulated $K^{0}_{S}$ mesons (top) and $\Lambda$ hyperons (bottom).}
\label{fig:p_theta_k0s_lambda}
\end{figure}

The simulation chain described in Ref.~\cite{NA61_pion} was used to correct raw yields for detector effects 
(i.e. geometrical acceptance, reconstruction efficiency, resolution of
$p$ and $\theta$ measurements), contributions from non-primary $V^{0}$ decays, 
and decay branching ratios. 
The correction factor was calculated according to Eq.~\ref{Eq:correction}. 
For this calculation information from the simulation about reconstructed 
and simulated particles of each type is needed. 
The number of reconstructed particles is obtained in the same way as for the real data as the result of the integration of the fitted signal function after background subtraction. The momentum versus polar angle distribution of simulated $K^{0}_{S}$ and $\Lambda$ from the employed VENUS~4.12 model~\cite{Venus412} is presented in Fig.~\ref{fig:p_theta_k0s_lambda}.

%\end{figure*}

\begin{figure*}[!h]
\begin{center}
\includegraphics[width=0.99\linewidth]{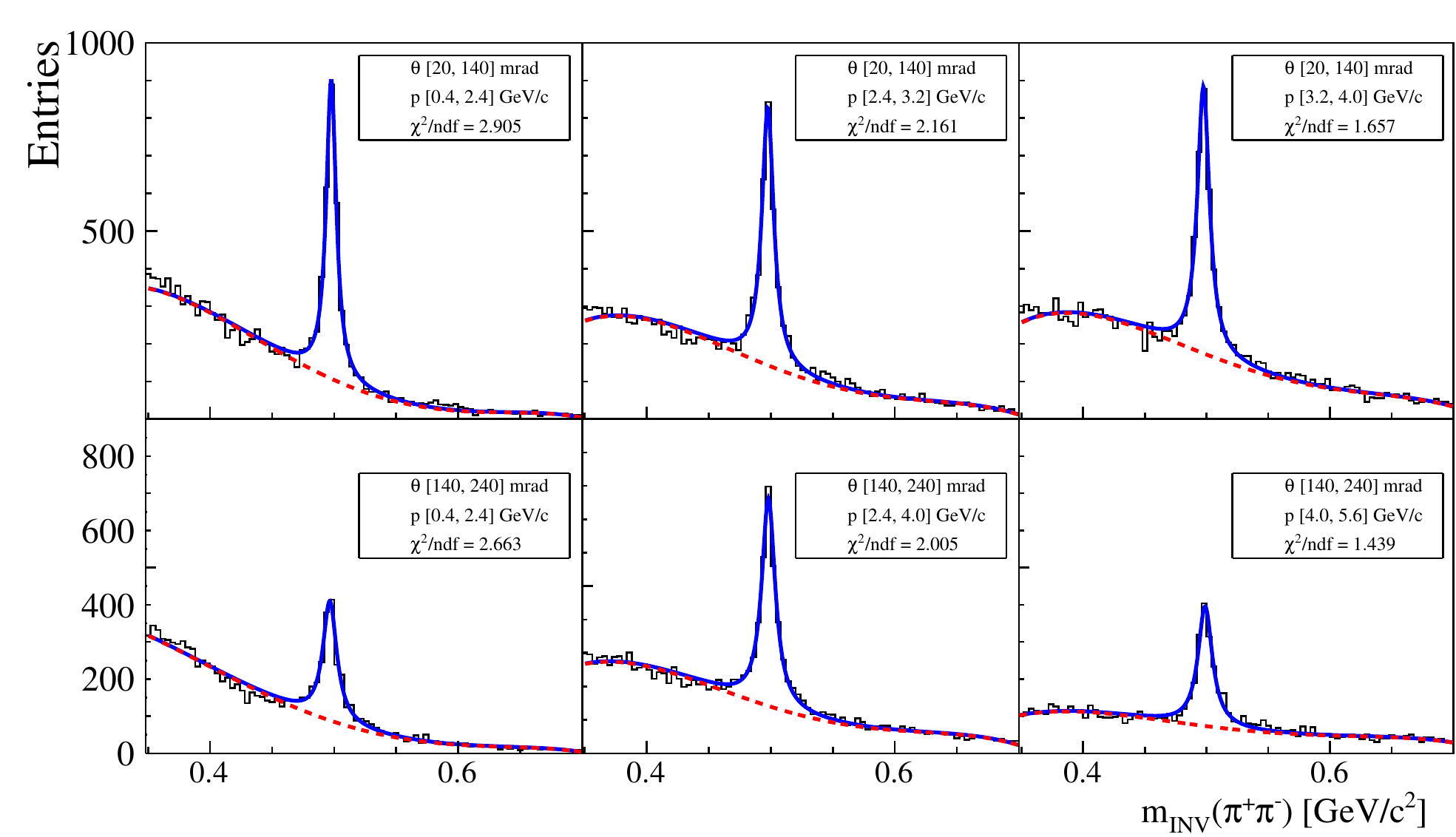}
\end{center}
  \caption{The  invariant mass distributions for simulated $K^{0}_{S}$ candidates 
in selected \{$p$, $\theta$\} bins.}
\label{fig:K0S_minv_mc}
\end{figure*}

\begin{figure*}[h!]
\begin{center}
\includegraphics[width=0.99\linewidth]{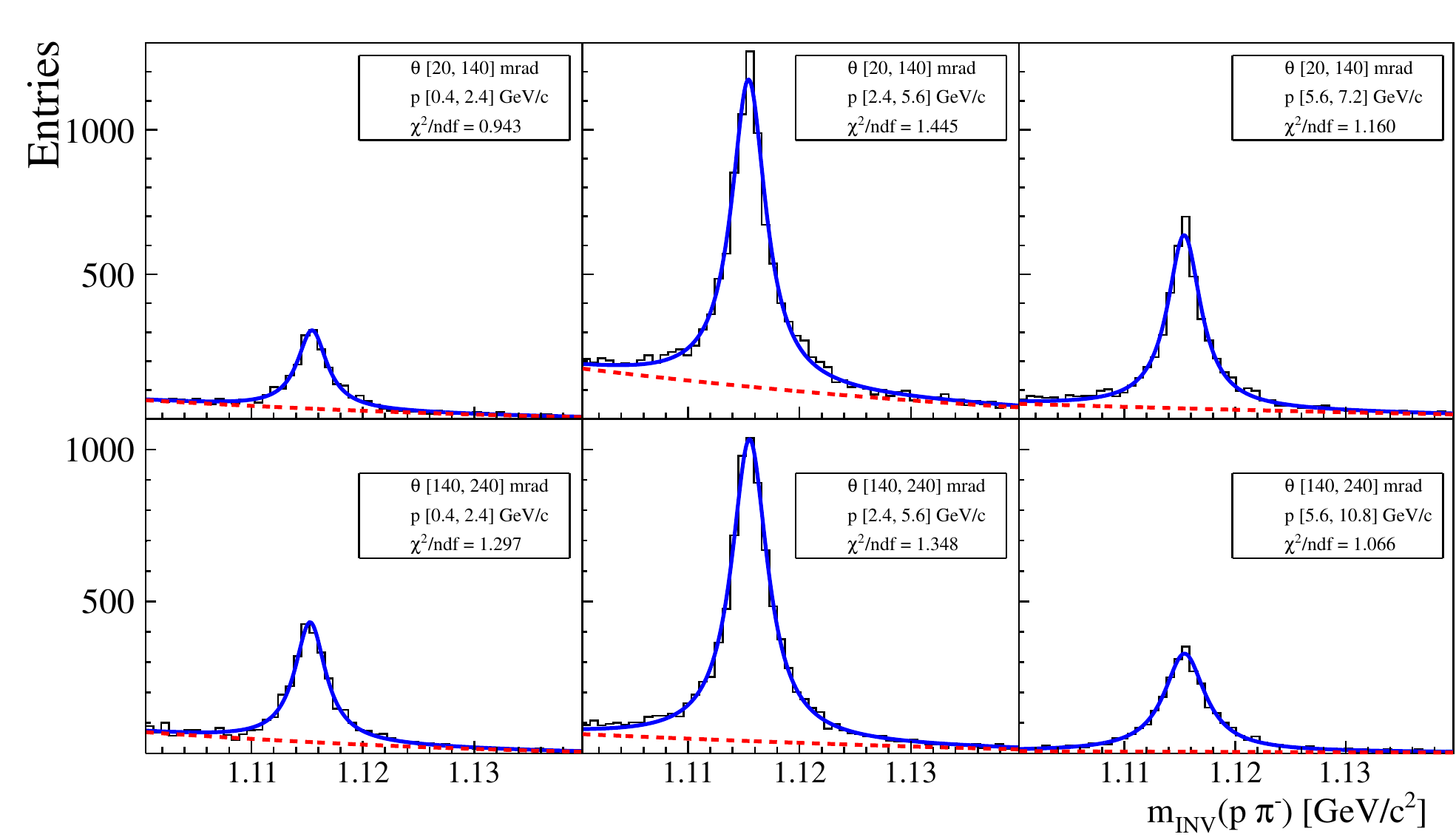}
\end{center}
  \caption{The invariant mass distributions for simulated $\Lambda$ candidates 
in selected \{$p$, $\theta$\} bins.}
\label{fig:Lambda_minv_mc}
\end{figure*}

The  invariant mass distributions for  $K^{0}_{S}$ and $\Lambda$  candidates 
in selected \{$p$, $\theta$\} bins 
are presented in Figs.~\ref{fig:K0S_minv_mc} and~\ref{fig:Lambda_minv_mc},
respectively.

\begin{figure}[!ht]
%\begin{figure*}[!h]
\begin{center}
\includegraphics[width=0.49\linewidth]{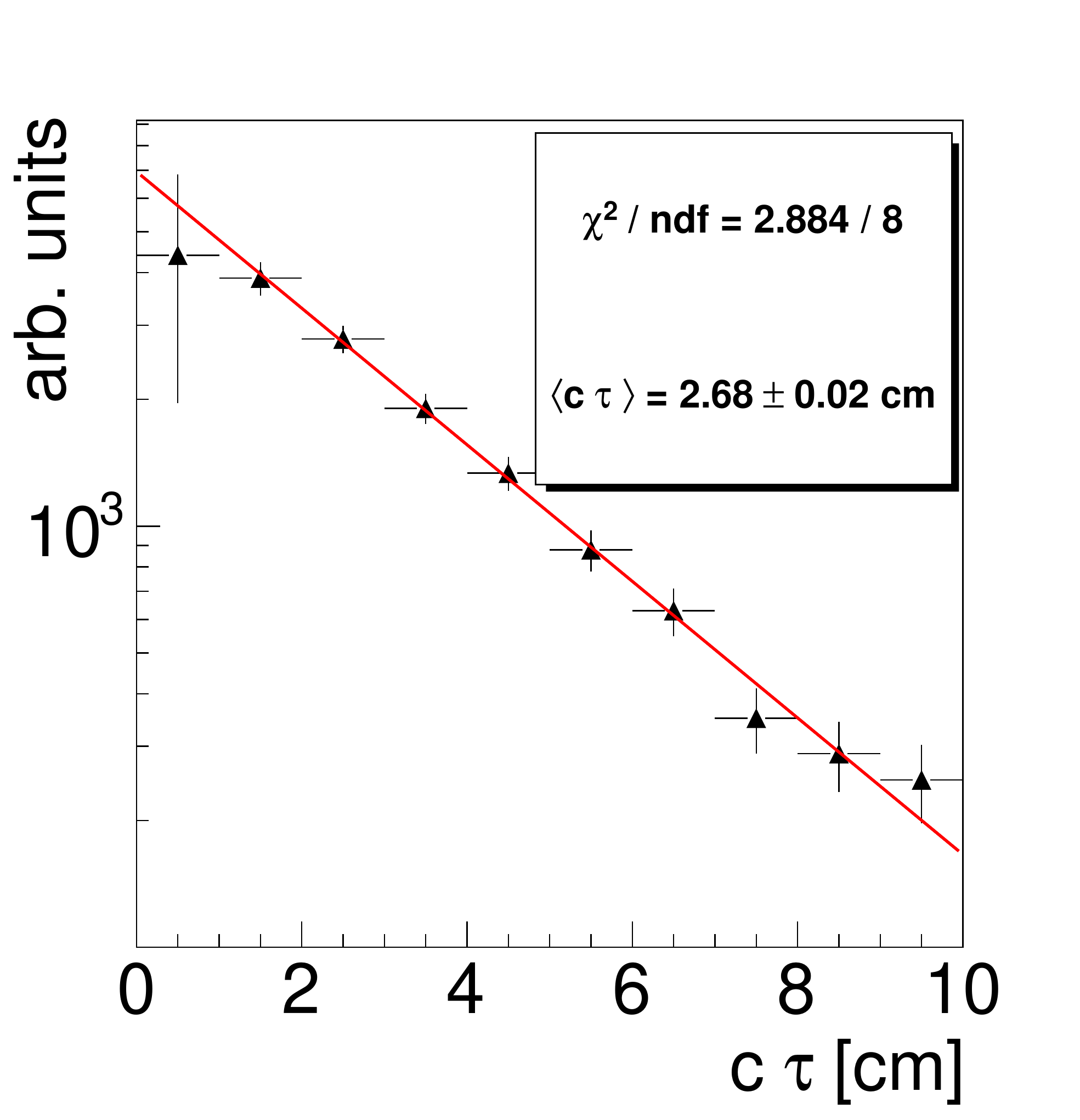}
\includegraphics[width=0.49\linewidth]{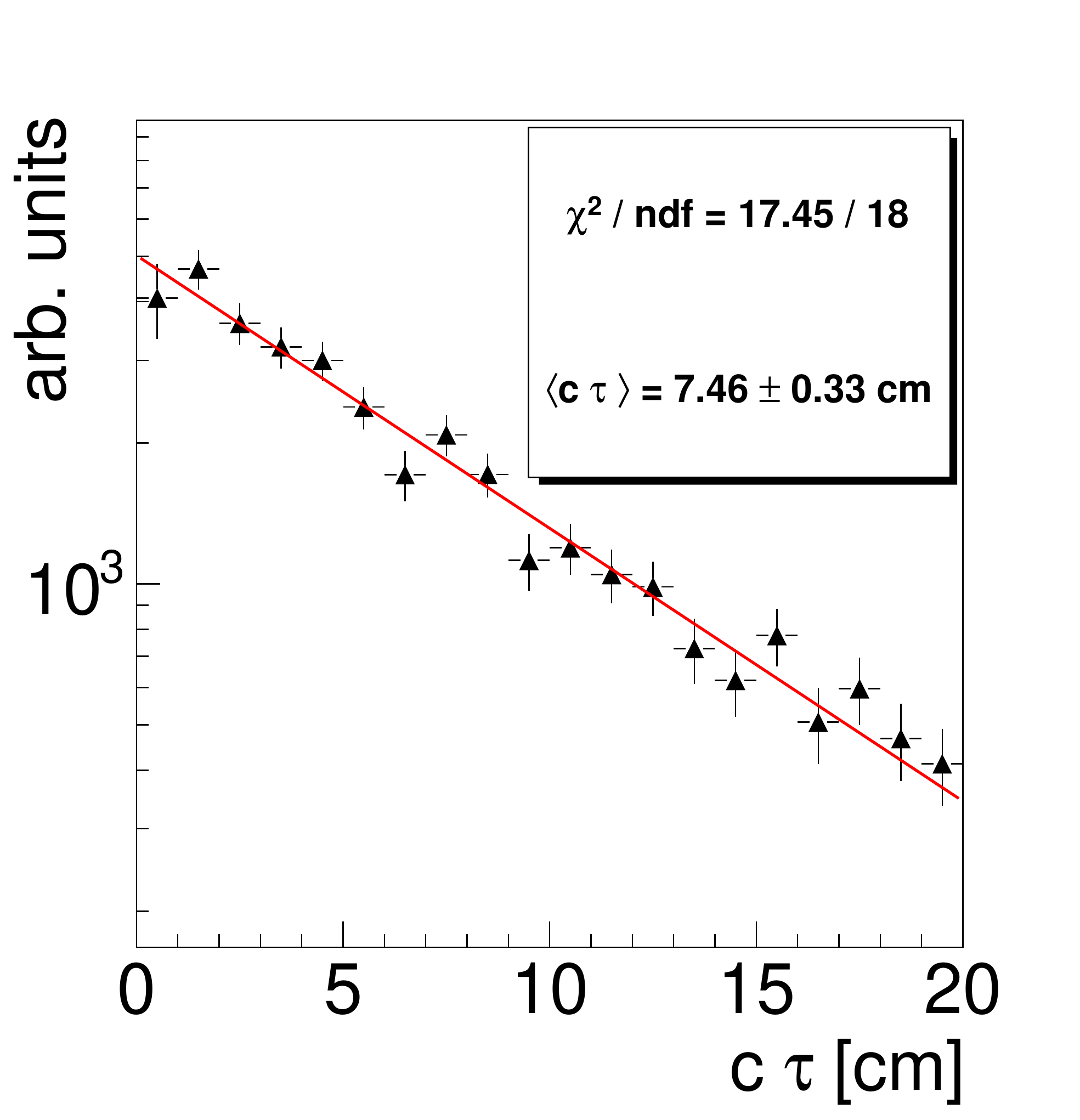}
\end{center}
  \caption{The proper decay length (c$\tau$) distributions for $K^{0}_{S}$ mesons (top) and $\Lambda$ hyperons (bottom).}
\label{fig:lifetime}
\end{figure}

The correction factors $\gamma$ for $K^{0}_{S}$ and $\Lambda$  
(see Eq.~\ref{Eq:correction_2}) were evaluated in bins of momentum and polar angle variables and are presented in Table \ref{tab:K0S_mc} and Table \ref{tab:Lambda_mc}, respectively. 
The correction factor $\eta$ is bin independent 
and was estimated to be equal to 0.8083 $\pm$ 0.0005. 
\begin{center}
\begin {table}[!hb]
\caption{
\label{tab:K0S_mc}
The number of simulated $K^{0}_{S}$ ($\mathrm n_\mathrm{sim}$), number of $K^{0}_{S}$ found  by the fitting procedure ($\mathrm n_\mathrm{rec}$), the $\gamma$-correction factor, 
and their errors in selected momentum and polar angle intervals.}
\begin{tabular}{l l|l l|l|l|l|l|l|l}
\hline
$\theta_\mathrm{low}$ & $\theta_\mathrm{up}$ & $\mathrm p_\mathrm{low}$ & $\mathrm p_\mathrm{up}$ & $\mathrm n_\mathrm{sim}$ & $\mathrm \Delta \mathrm n_\mathrm{sim}$ & 
$\mathrm n_\mathrm{rec}$ & $\Delta \mathrm n_\mathrm{rec}$ & $\gamma$ & $\Delta$ $\gamma$ \\
\multicolumn{2}{c|}{(mrad)} & \multicolumn{2}{c|}{(GeV/c)} & & & & &   \\
\hline \hline
  0& 20	& 0.4   &7.2    &10908   &104  &1067 & 37& 0.0979 & 0.0035 \\
\hline
  20& 140	& 0.4   &2.4    &58087   &241  &2417 & 55& 0.0416 & 0.001 \\
  & 	        & 2.4   &3.2    &32905   &181  &2472 & 59& 0.0751 & 0.0018 \\
  &  	        & 3.2   &4.0    &31544   &178  &2624 & 62& 0.0831 & 0.0020 \\
  &     	& 4.0   &4.8    &28274  &168  &2574 & 62& 0.0910 & 0.0022 \\
  &    	& 4.8   &5.6    &24330   &156  &2241 & 57& 0.0921 & 0.0024 \\
  &     	& 5.6   &7.2    &36936   &192  &3345 & 71& 0.0905 & 0.0020 \\
  &     	& 7.2   &8.8    &24172   &155  &2325 & 60& 0.0962 & 0.0025 \\
  &     	& 8.8   &10.8    &18054   &134  &1726 & 52& 0.0956 & 0.0029 \\
  &     	& 10.8   &20.0    &16742   &129  &1521 & 54& 0.0909 & 0.0033 \\
\hline
140  & 240  & 0.4   &2.4    &69280   &263  &1359 & 45& 0.0196 & 0.0006 \\
  &     	   & 2.4   &4.0    &31443   &177  &2548 & 61& 0.0810 & 0.0020 \\
  &     	   & 4.0   &5.6    &10373   &102  &1417 & 46& 0.1366 & 0.0046 \\
  &     	   & 5.6   &10.8    &4250   &65  &661 & 34& 0.1557 & 0.0083 \\
\hline
240  &  420  & 0.4   &5.6    & 119075  &345  &2209 & 61& 0.0185 & 0.0005 \\
\hline
 \end{tabular}
\end {table}
\end{center}
\begin{center}
\begin {table}[!hb]
\caption{
\label{tab:Lambda_mc}
The number of simulated $\Lambda$ ($\mathrm n_\mathrm{sim}$), number of $\Lambda$ found by the fitting procedure ($\mathrm n_\mathrm{rec}$), $\gamma$ correction factor, and their errors in selected momentum and polar angle intervals.}
\begin{tabular}{l l|l l|l|l|l|l|l|l}
\hline
$\mathrm \theta_\mathrm{low}$ & $\mathrm \theta_\mathrm{up}$ & $\mathrm p_\mathrm{low}$ & $\mathrm p_\mathrm{up}$ & $\mathrm n_\mathrm{sim}$ & $\Delta \mathrm n_\mathrm{sim}$ & 
$\mathrm n_\mathrm{rec}$ & $\Delta \mathrm n_\mathrm{rec}$ & $\gamma$ & $\Delta$ $\gamma$ \\
\multicolumn{2}{c|}{(mrad)} & \multicolumn{2}{c|}{(GeV/c)} & & & & &   \\
\hline \hline
  20& 140	& 0.4   &2.4    &38316   &196  &1557 & 43& 0.041 & 0.001 \\
  & 	        & 2.4   &5.6    &62777   &250  &7081 & 90& 0.113 & 0.001 \\
  &  	        & 5.6   &7.2    &30357   &174  &3896 & 64& 0.128 & 0.002 \\
  &     	& 7.2   &10.8    &79015 &281  &9122 & 95& 0.115 & 0.001 \\
  &     	& 10.8   &20.0    &121755   &349  &10883 & 102& 0.089 & 0.001 \\
\hline
140  & 240  & 0.4   &2.4    &61866   &249  &2377 & 52& 0.038 & 0.001 \\
  &     	   & 2.4   &5.6    &58957   &243  &6929 & 84& 0.117 & 0.001 \\
  &     	   & 5.6   &10.8    &15894   &126  &2595 & 51& 0.163 & 0.003 \\
\hline
 \end{tabular}
\end {table}
\end{center}  
The correction procedure was checked by evaluating the lifetime of $K^{0}_{S}$ and $\Lambda$ from the corrected distribution of distances between the production and decay vertices. 
The proper decay length (c$\tau$) distributions for $K^{0}_{S}$ mesons and $\Lambda$ hyperons are shown in Fig.~\ref{fig:lifetime}.

%\end{figure*}

The obtained average proper decay lengths of 2.68~$\pm$~0.02~cm for $K^{0}_{S}$ and 7.46~$\pm$~0.33~cm for $\Lambda$ are consistent with the world averages. 

%The resulting values of $(0.89 \pm 0.05)$~ps for $K^{0}_{S}$ and $(2.49 \pm 0.11)$~ps for $\Lambda$ are %consistent with the world averages. 

The contribution of non-target interactions has small impact on the final results. 
The ratio B was found to be equal to  2.198 $\pm$ 0.027. 
After all cuts there were no $V^{0}$ candidates found in the data recorded 
with target removed. 
Therefore, the subtraction procedure for out of target interactions 
changes only the number of events used for the 
normalization of spectra. The values of $\mathrm N^\mathrm{I}$ and $\mathrm N^{\mathrm I}$-$\mathrm B \mathrm N^{\mathrm R}$ were 
276481 and  276421, respectively.

%____________________________________________________________________________ 

\subsection{Systematic errors}
\label{Sec:Sys}
The main sources of the systematic uncertainty are as follows. 

\begin{enumerate}[(i)]
\item{The uncertainty connected with the track cuts. 
The contribution of this source was studied by varying all standard track cuts (see Sec.~\ref{Sec:binning}). In particular the minimum distance between the primary vertex and the $V^0$ decay point was varied from 3~cm to 6~cm. The invariant mass window was changed to [0.4, 0.65] GeV/c$^{2}$ and to [1.09, 1.16] GeV/c$^{2}$ for $K^{0}_{S}$ and $\Lambda$, respectively. The accepted region for the cos~$\epsilon$ was reduced to [-0.8, 0.6] for $K^{0}_{S}$ and to [-0.7,0.7] for $\Lambda$. The cuts in the Armenteros-Podolanski plots were varied by about 10\%.
 } 

\item{The uncertainty connected with the background function used 
in the fitting procedure. 
Apart from the standard $4^{th}$ order polynomial 
a set of different background functions was 
tried ($3^{rd}$ order polynomial, $6^{th}$ order Chebyshev polynomial, 
Argus function).}

\item{The uncertainty connected with the fitting procedure. 
Due to the low statistics of the data, fit results depend on the fit strategy and
the limits set for the parameters. 
The most prominent effect was observed when the width of the signal function was varied.
The initial values of the position and the width of the signal were taken
from the large statistic MC simulations and the variation within $\pm$10$\cdot$$\Delta F_{MC}$ was used for the systematic error studies.
 }    

\item{ The uncertainty connected with the inaccurate description of the reconstructed primary vertex distribution in the Monte Carlo.}

\item{ The uncertainty connected with the inaccuracy of the geometrical
acceptance obtained from Monte Carlo simulation, reconstruction efficiency, 
and different algorithms used for track merging.}

\item{The uncertainty connected with the \{$p$, $\theta$\} bin size
due to imperfect modeling of $K^0_S$ and $\Lambda$ spectra in the Monte Carlo. 
This effect was only studied in the $K^{0}_{S}$ case by looking at 
the distributions in different Monte Carlo generators.}

\end{enumerate}

The systematic errors connected with uncertainties discussed in points (i) 
and (iv) were found to be almost momentum independent. 
The uncertainties discussed in  point (v) depend strongly 
on the selected \{$p$, $\theta$\} bin. The systematic errors were estimated following the procedure presented 
in Refs.~\cite{NA61_pion, NA61_Kplus}. 

The final systematic errors, calculated as the sum in quadrature of errors connected with the uncertainties discussed above, were found to be about 13.5-20.0\% for $K^{0}_{S}$ and $\Lambda$ depending on the emission angle interval. The largest contributions always come from 
the uncertainty connected with the fitting approach (iii) and the uncertainty connected with the fit of the background (ii). Typical systematic errors connected with these sources
were about 8-12\% (iii) and about 7-8\% (ii). 
The second most important contribution is related to the track cuts (i) and it is typically  4-5\%. 
In addition important contribution comes from an inaccurate  agreement 
between the data and Monte Carlo distributions of the fitted $z$ coordinate of the
primary  vertex (iv). The systematic error connected with this effect is about 5\%.  
The contributions connected with geometrical acceptance, reconstruction efficiency, 
and different reconstruction algorithms (v) are 3-4\%, 2\% and 2\%, respectively. 
The systematic error connected with bin size (vi) was found to be about 6\% in  highly 
populated bins and about 12\% in  low statistic bins.
%____________________________________________________________________________ 

\section{Results}
\label{Sec:results}
\begin{figure*}[!ht]
\begin{center}
\includegraphics[width=0.45\linewidth]{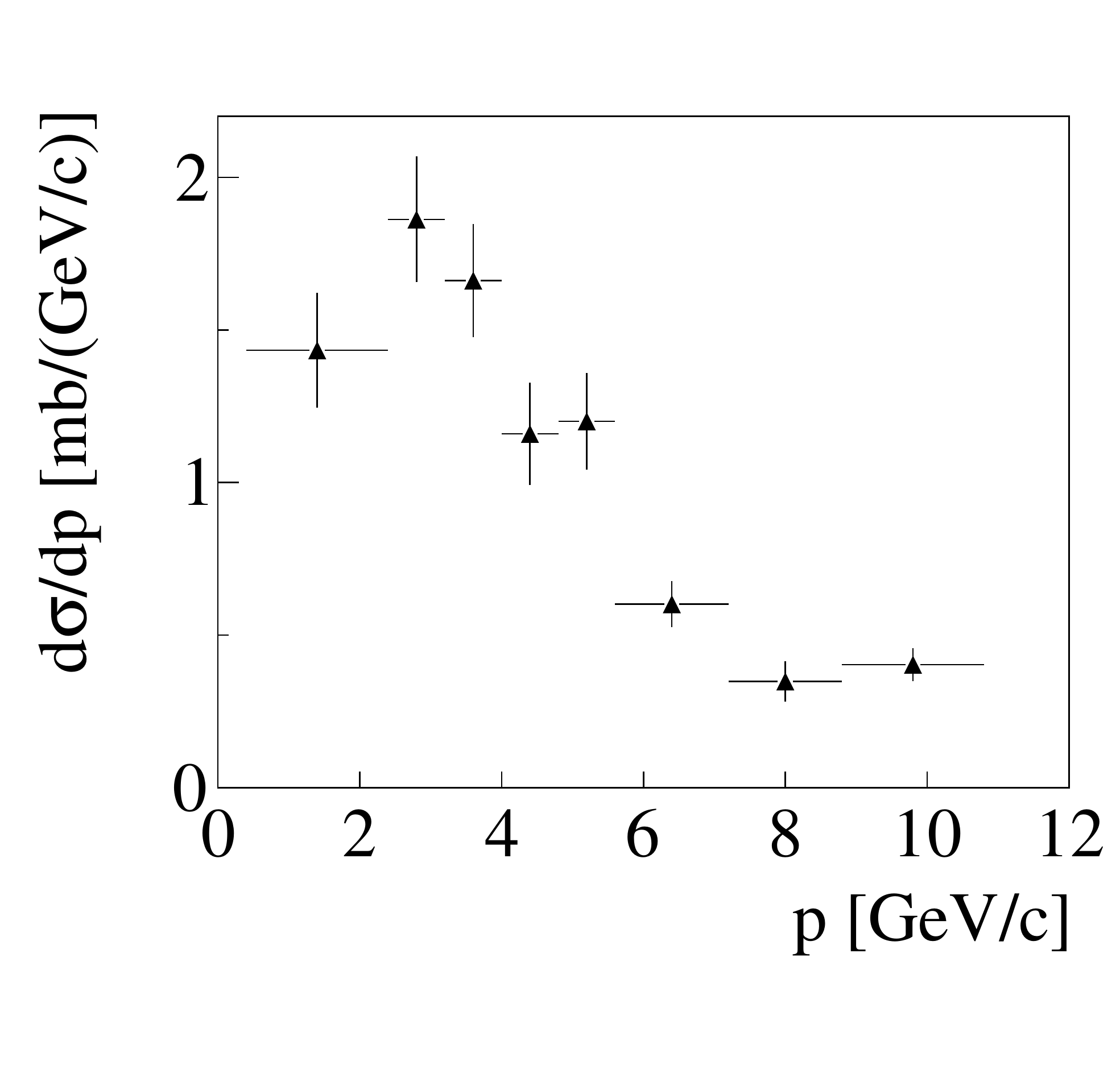}
\includegraphics[width=0.45\linewidth]{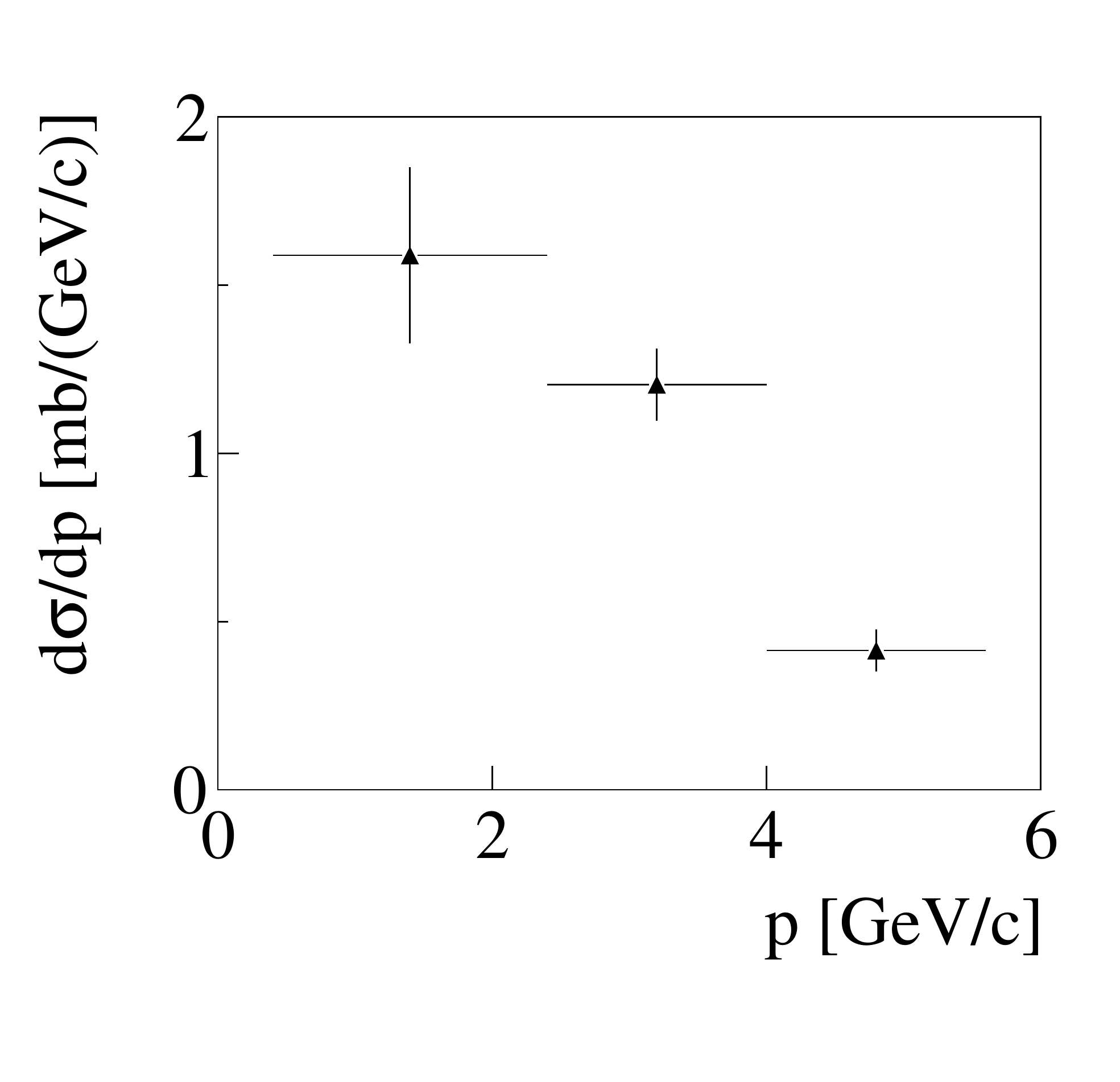}
\end{center}
 \caption{$K^{0}_{S}$ production cross sections in two polar angle intervals: [20, 140] mrad (left) and [140, 240] mrad (right). Only statistical errors are shown. 
The normalization uncertainty is not included. }
\label{fig:K0S_final_dsigmadp}
\end{figure*}

\begin{figure*}[ht!]
\begin{center}
\includegraphics[width=0.45\linewidth]{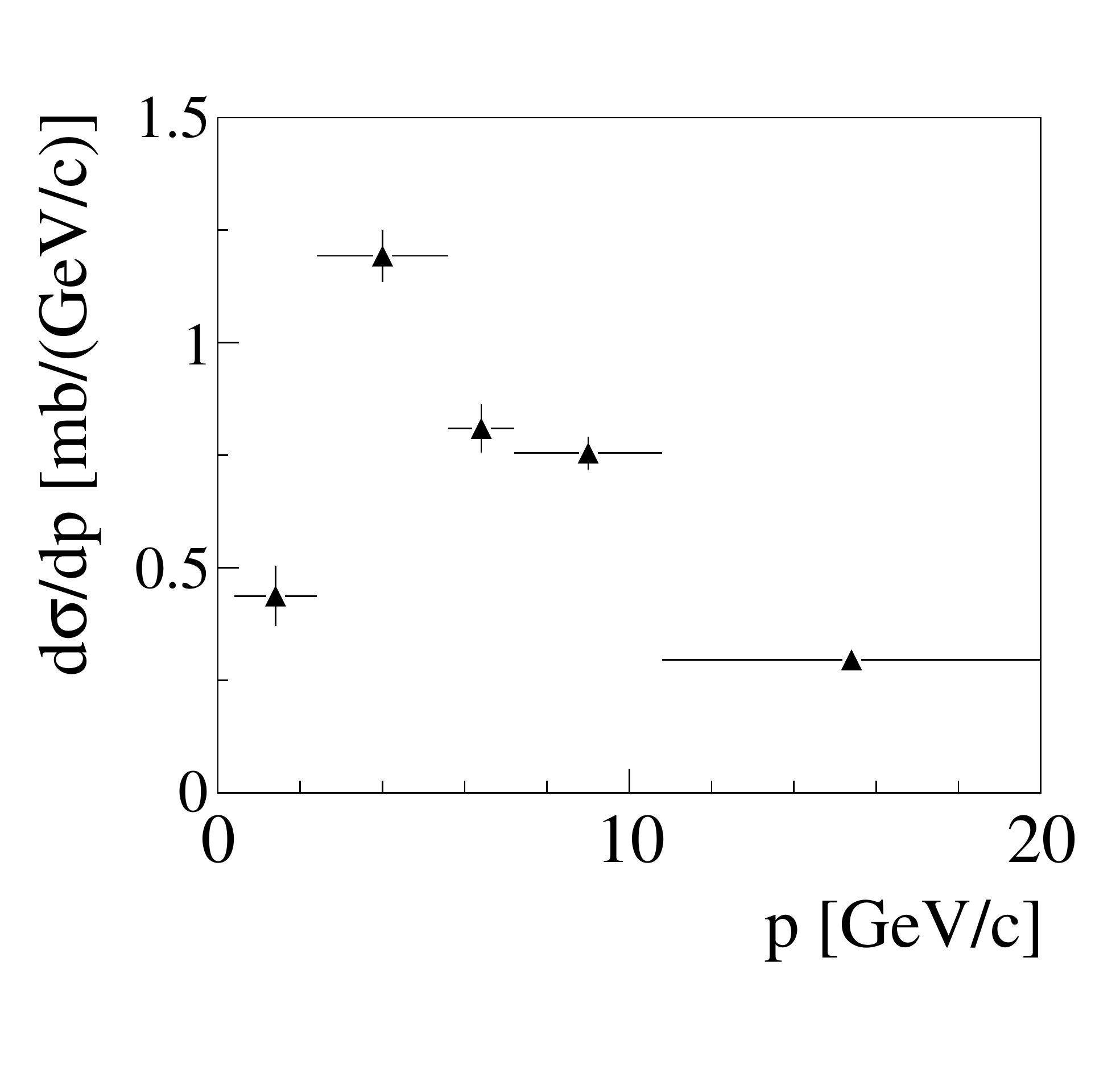}
\includegraphics[width=0.45\linewidth]{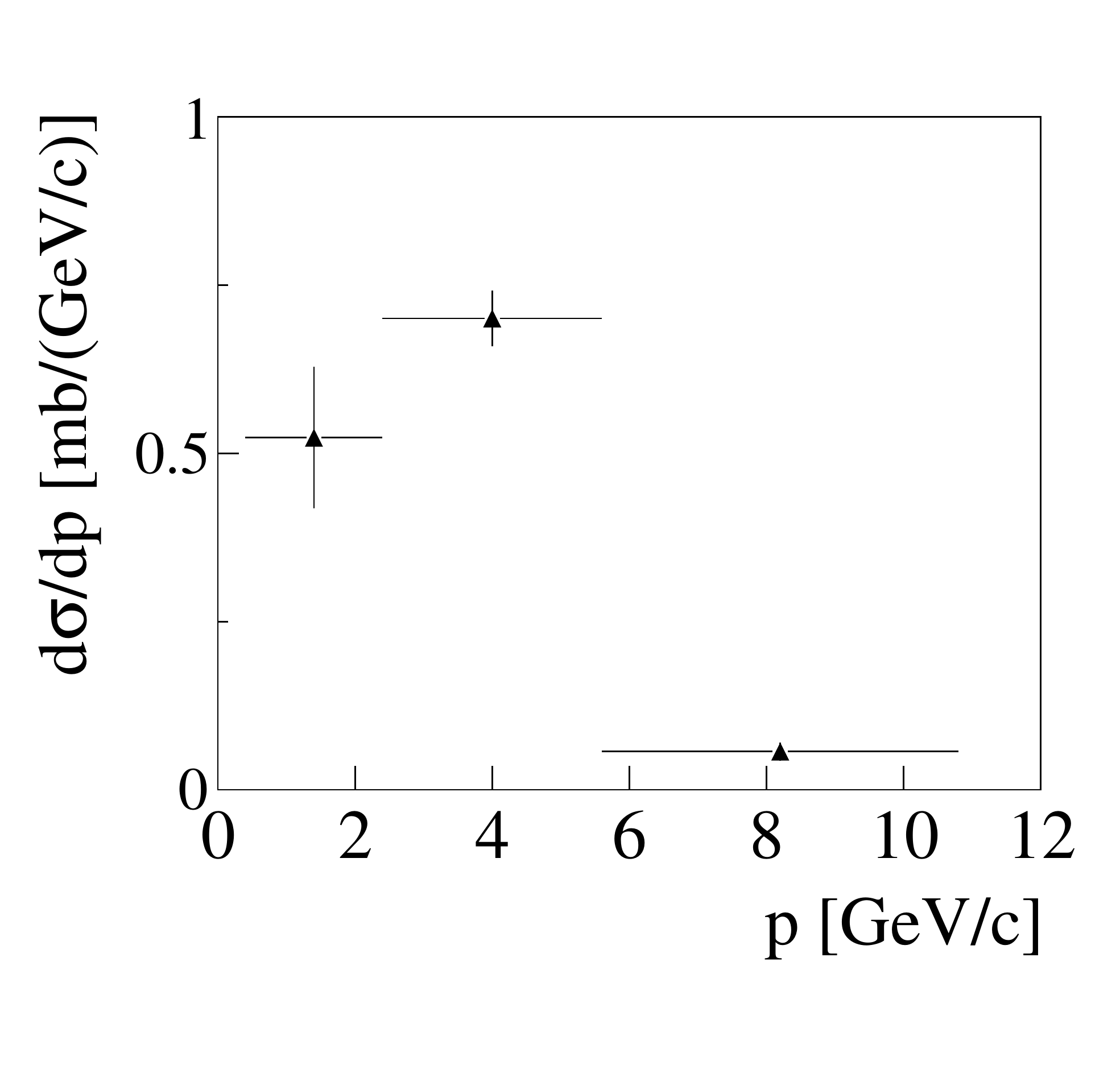}
\end{center}
  \caption{$\Lambda$ production cross sections in two polar angle intervals: 
[20, 140] mrad (left), [140, 240] mrad (right). 
Only statistical errors are shown. 
The normalization uncertainty is not included.}
\label{fig:Lambda_final_dsigmadp}
\end{figure*}

Differential inclusive cross sections were derived following 
the procedure described in Sec.~\ref{Sec:analysis}. 
They are tabulated in Tables~\ref{tab:K0S_final_results} 
and \ref{tab:Lambda_final_results}
as well as plotted in polar angle slices in Figs.~\ref{fig:K0S_final_dsigmadp} 
and \ref{fig:Lambda_final_dsigmadp}
for $K^{0}_{S}$ and $\Lambda$, respectively.

%\begin{figure}[!ht]
%\end{figure}

\begin{center}
\begin {table}[h!]
\caption{
\label{tab:K0S_final_results}
The NA61/SHINE results for the differential $K^{0}_{S}$ production cross section in the laboratory system for p+C interactions at 31 GeV/c.}
\begin{tabular}{l l|l l|l l l l}
\hline
$\mathrm \theta_\mathrm{low}$ & 
$\mathrm \theta_\mathrm{up}$  & 
$\mathrm p_\mathrm{low}$        & 
$\mathrm p_\mathrm{up}$         & 
d$\sigma_\mathrm{K^{0}_{S}}$ $/$ dp &
$\Delta_\mathrm{stat}$  &
$\Delta_\mathrm{stat}$ &
$\Delta_\mathrm{sys}$  \\
\multicolumn{2}{c|}{ (mrad)} &
\multicolumn{2}{c|}{ (GeV$/$c)} &
\multicolumn{2}{c}{ (mb$/$GeV$/$c)} &
[\%] &
[\%] \\
\hline \hline
 0& 20	& 0.4   & 7.2   &0.061    &0.010   &16.8  & 18.4   \\
\hline
  20& 140	& 0.4   &2.4    &1.434   &0.188  &13.1 & 17.8  \\
  & 	        & 2.4   &3.2    &1.862   &0.206  &11.0 &  15.5 \\
  &  	        & 3.2   &4.0    &1.662   &0.186  &11.2 &  14.4 \\
  &     	& 4.0   &4.8    &1.159  &0.168  &14.5 &  14.4\\
  &    	               & 4.8   &5.6    &1.200   &0.158  &13.1 & 14.4    \\
  &     	& 5.6   &7.2    &0.601   &0.075  &12.5 &  14.4 \\
  &     	& 7.2   &8.8    &0.348   &0.066  &18.9 &  14.4 \\
  &     	& 8.8   &10.8    &0.403   &0.054  &13.4 & 16.3  \\
  &     	& 10.8   &20.0    &0.082   &0.014  &17.0 &  18.6\\
\hline
140  & 240  & 0.4   &2.4    &1.589   &0.26  &16.5 &  19.5 \\
  &     	   & 2.4   &4.0    &1.204   &0.139  &8.9 &  14.4 \\
  &     	   & 4.0   &5.6    &0.414  & 0.063 &15.2 &  14.4 \\
  &     	   & 5.6   &10.8    &0.073  &0.011  &14.6 &  16.9 \\
\hline
240  &  420  & 0.4   &5.6    & 0.938  &  0.182 & 19.4 & 20.6 \\
\hline
 \end{tabular}
\end {table}
\end{center}
\begin{center}
\begin {table}[h!]
\caption{
\label{tab:Lambda_final_results}
The NA61/SHINE results for differential $\Lambda$ production cross section in the laboratory system for p+C interactions at 31 GeV/c.}
\begin{tabular}{l l|l l|l l l l}
\hline
$\mathrm \theta_\mathrm{low}$ & 
$\mathrm \theta_\mathrm{up}$  & 
$\mathrm p_\mathrm{low}$        & 
$\mathrm p_\mathrm{up}$         & 
d$\sigma_\mathrm{\Lambda}$ $/$ dp &
$\Delta_\mathrm{stat}$  &
 $\Delta_\mathrm{stat}$ &
$\Delta_\mathrm{sys}$  \\
\multicolumn{2}{c|}{ (mrad)} &
\multicolumn{2}{c|}{ (GeV$/$c)} &
\multicolumn{2}{c}{ (mb$/$GeV$/$c)} &
[\%] &
[\%] \\
\hline \hline
  20& 140	& 0.4   &2.4    & 0.437   & 0.067   & 15.3 & 16.8  \\
  & 	        & 2.4   &5.6    & 1.193   & 0.057  & 4.8 &  16.1  \\
  &  	        & 5.6   &7.2    & 0.809 & 0.054   & 6.7 &  15.9 \\
  &     	& 7.2   &10.8  & 0.755  & 0.036   & 4.8  & 16.6 \\
  &    	         & 10.8   &20.0& 0.295  & 0.016  & 5.3 & 16.8   \\
 \hline
140  & 240  & 0.4   &2.4    & 0.523  & 0.105  & 20.1   & 16.6 \\
  &     	   & 2.4    &5.6    & 0.701  & 0.042  & 5.9    & 15.9 \\
  &     	   & 5.6   &10.8   & 0.057  & 0.013  & 23.7  & 16.8  \\
\hline
\end{tabular}
\end {table}
\end{center}

The mean multiplicity in production processes and the inclusive cross section for $K^{0}_{S}$ production were evaluated from the results obtained in momentum and polar angle intervals. Regions outside the geometrical acceptance were corrected according to the VENUS~4.12 model~\cite{Venus412}  which predicts 23.91\% of simulated $K^{0}_{S}$ to be outside of our \{$p$, $\theta$\} bins. 
In the UrQMD~\cite{Urqmd131} model this number is 20.27\%. 

The spread of model predictions was used to estimate the systematic error of the extrapolation
to full phase space. \\
The final results with systematic errors are as follows. \\

	The $K^{0}_{S}$ mean multiplicity in production processes is\\
\centerline{$\langle n_{K^{0}_{S}}\rangle$ = 0.127 $\pm$ 0.005 (stat) $\pm$ 0.022 (sys)~.}\\

	The inclusive cross section for $K^{0}_{S}$ production is\\
\centerline{$\sigma_\mathrm{K^{0}_{S}}$ = 29.0 $\pm$ 1.6 (stat) $\pm$ 5.0 (sys) [mb]~.}\\

Due to different phase-space distributions of $\Lambda$ and $K^{0}_{S}$ 
a model-dependent 
correction for a unmeasured yield of 
$\Lambda$ hyperons is about two times larger than the one
for $K^{0}_{S}$ mesons. Therefore a reliable extrapolation of 
the $\Lambda$ hyperon yield to the full phase
was not possible.

\begin{figure}[hbt!]
\begin{center}
\includegraphics[width=0.9\linewidth]{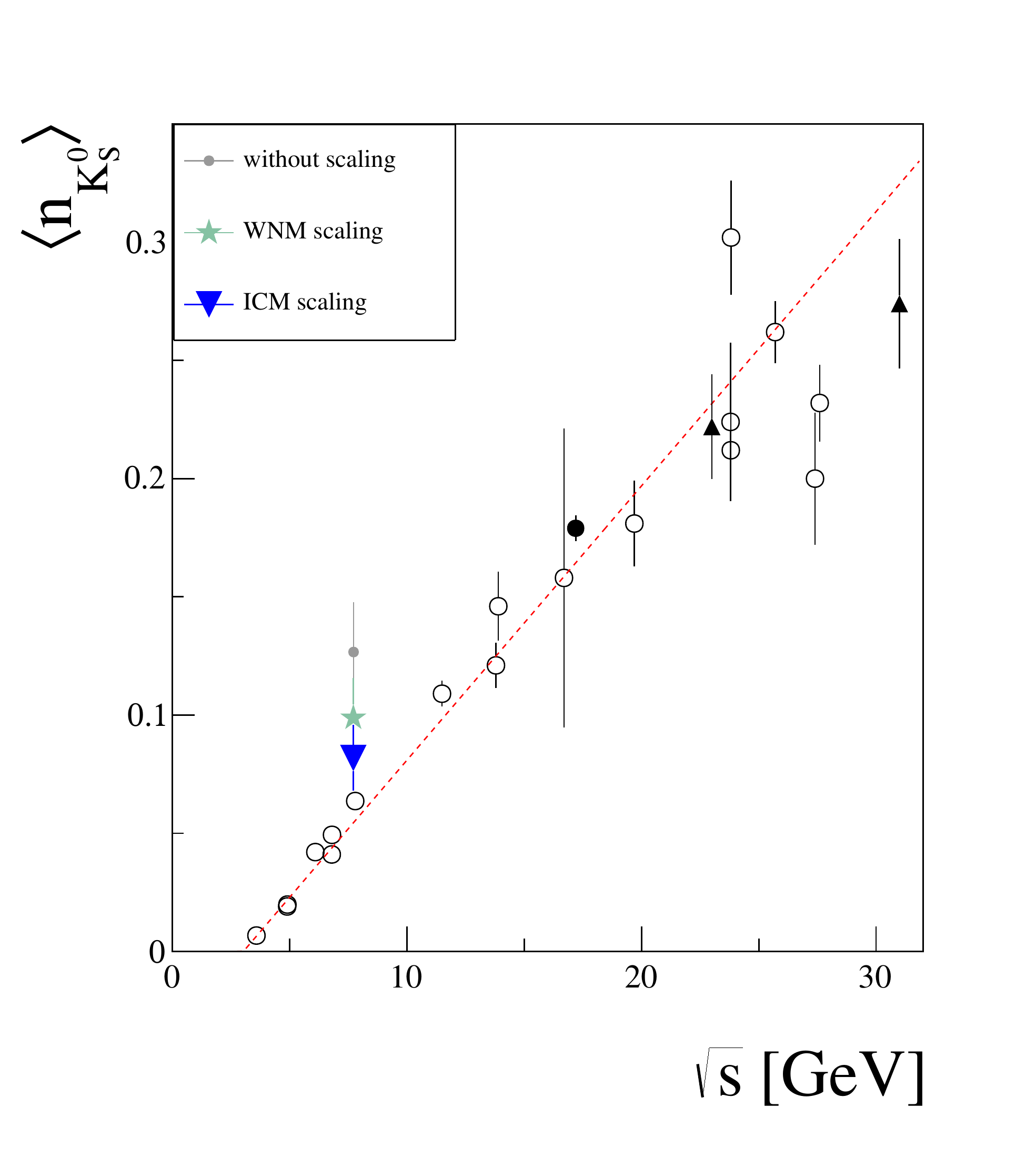}
\end{center}
  \caption{Mean multiplicity of $K^{0}_{S}$ mesons as a function of $\sqrt{s}$. 
Open black points: p+p results from the compilation ~\cite{pp_compilation}. 
Full circle: the NA49 result on (($K^{+}$ + $K^{-}$)$/2$)~\cite{pp_compilation}. 
Triangles: the ISR measurements of (($K^{+}$ + $K^{-}$)$/2$) and the UA5 results~\cite{pp_compilation}. 
Grey circle: the result from this paper for p+C interactions. 
Green full star: the result from this paper after scaling according to the WNM model. 
Blue triangle: the result from this paper after scaling according to the ICM model. 
The dashed line is shown to guide the eye.}
\label{fig:K0S_mean_to_pp}
\end{figure}

The $K^{0}_{S}$ mean multiplicity in production processes from p+C interactions 
was compared with a compilation~\cite{pp_compilation} of 
total integrated $K^{0}_{S}$ yields from p+p interaction experiments. 
For this purpose the NA61/SHINE result was scaled according to the Wounded Nucleon Model (WNM) by a multiplicative scaling factor 2/(1+$\langle n_{W} \rangle$) and according to the Independent Collisions Model (ICM) by a multiplicative scaling factor 1/$\langle n_{W} \rangle$, 
where $\langle n_{W} \rangle$ is the average number of wounded nucleons inside the carbon nucleus. 
The value of $\langle n_{W} \rangle$ was found to be 
1.5240~$\pm$~0.0011 using the GLauber Initial-State Simulation AND mOre (GLISSANDO) model calculation~\cite{glissando}. 
The comparison of our result with the compilation of existing p+p results is shown in Fig.~\ref{fig:K0S_mean_to_pp}. 
Better agreement is observed between the scaled NA61/SHINE result and the measurement for p+p interactions when the ICM scaling is used. The larger value of the kaon  multiplicity (0.127 $\pm$ 0.005) as compared to the WNM model
($0.063 \times 1.26 = 0.0793$) can be regarded as an indication of enhanced  strange particle production in p+C collisions  at 31~GeV/c. However, a definitive conclusion is not possible due to the systematic uncertainties of our results.  

%____________________________________________________________________________ 

\section{Comparison with model predictions}
\label{Sec:MC}

\begin{figure*}[ht!]
\begin{center}
\includegraphics[width=0.49\linewidth]{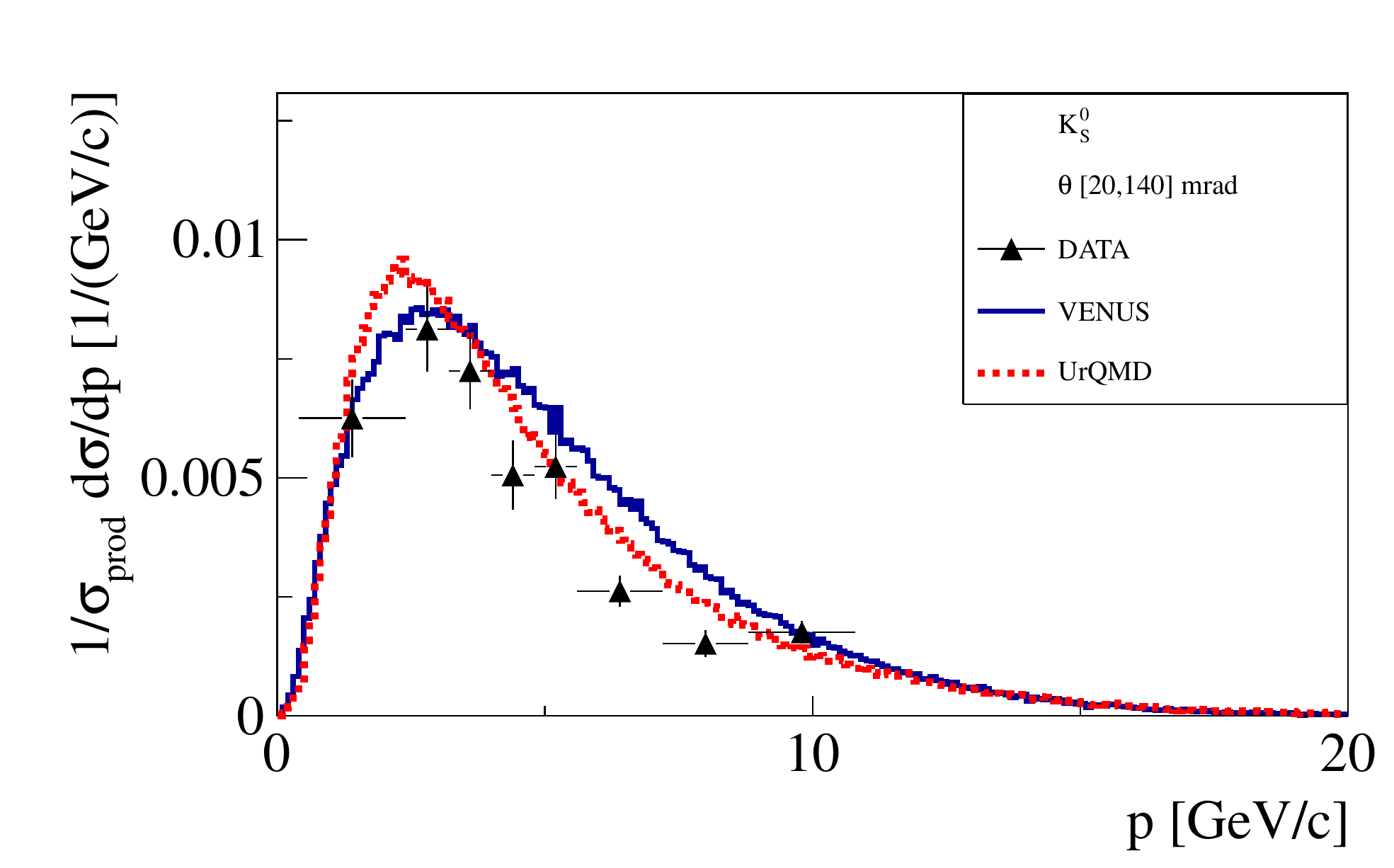}
\includegraphics[width=0.49\linewidth]{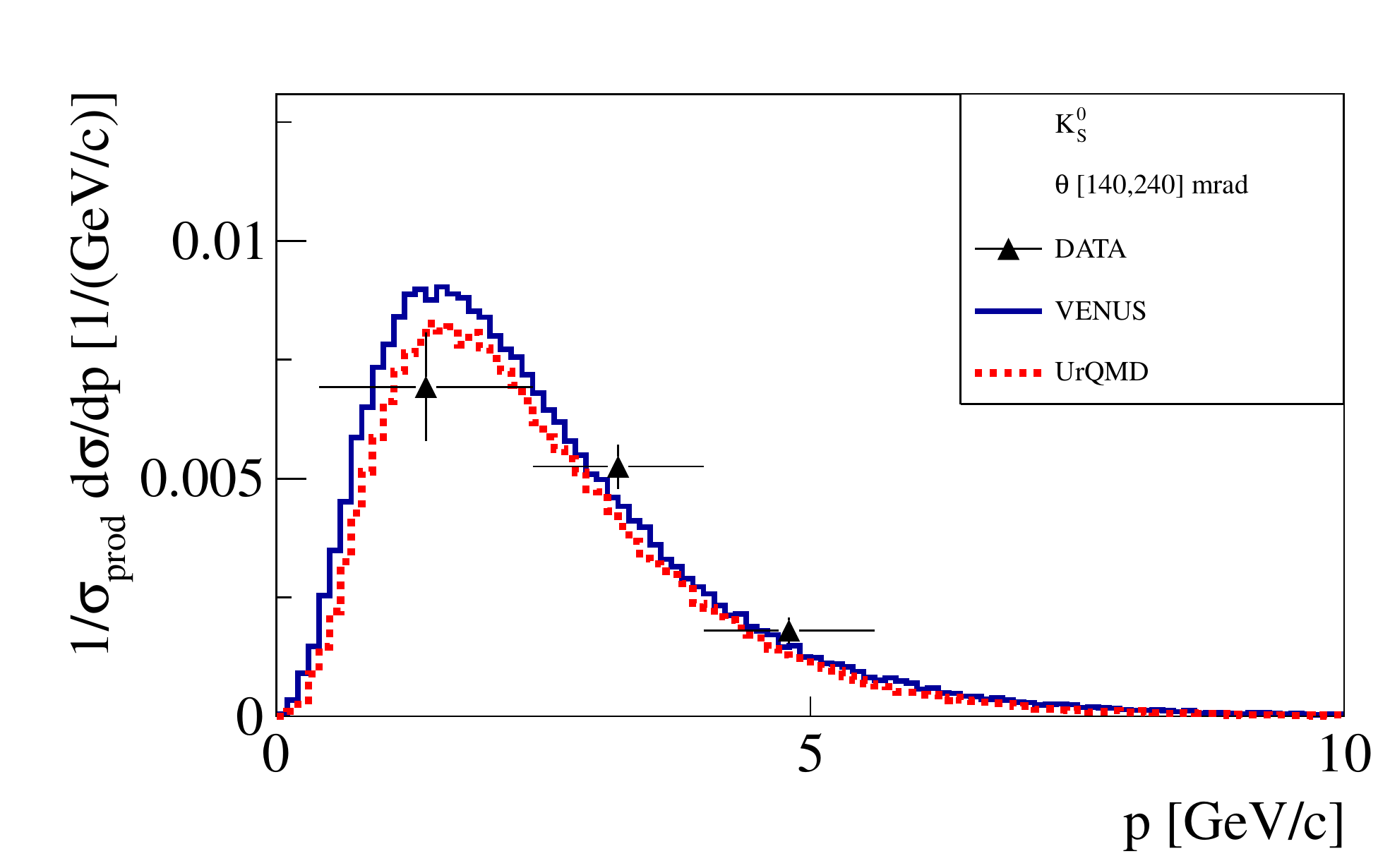}
\end{center}
  \caption{Mean multiplicity of $K^{0}_{S}$ mesons in production processes in polar angle intervals: [20, 140] mrad (left), [140, 240] mrad (right). Two hadron production model predictions are superimposed. Only statistical errors are shown.}
\label{fig:K0S_dn_dp}
\end{figure*}

The $K^{0}_{S}$ results in momentum and polar angle variables normalized to mean particle multiplicity in production processes are shown in Fig.~\ref{fig:K0S_dn_dp} with predictions from the hadron production models VENUS and UrQMD superimposed.

\begin{figure}[ht!]
%\begin{figure*}[ht!]
\begin{center}
\includegraphics[width=0.67\linewidth]{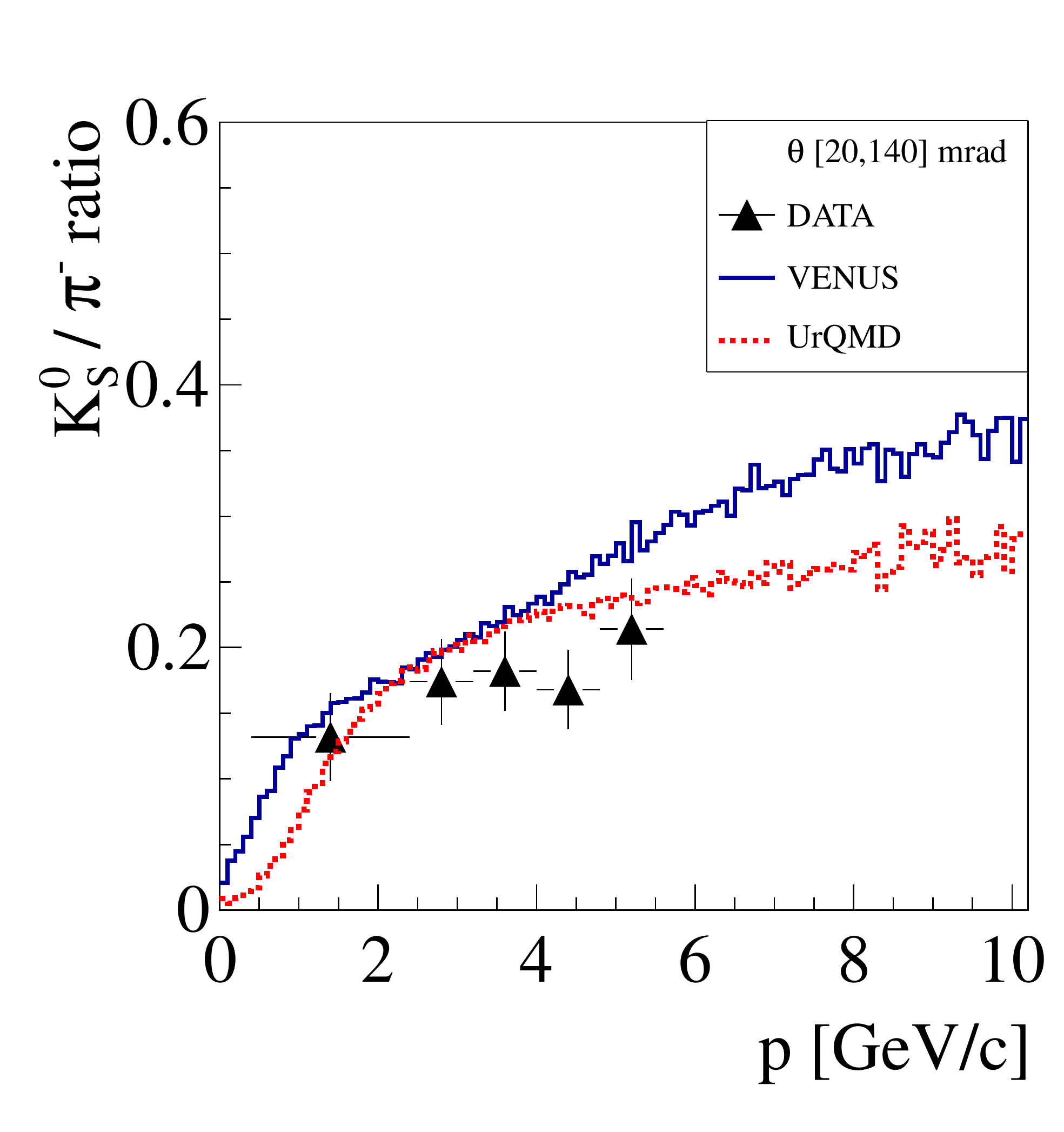}
\includegraphics[width=0.67\linewidth]{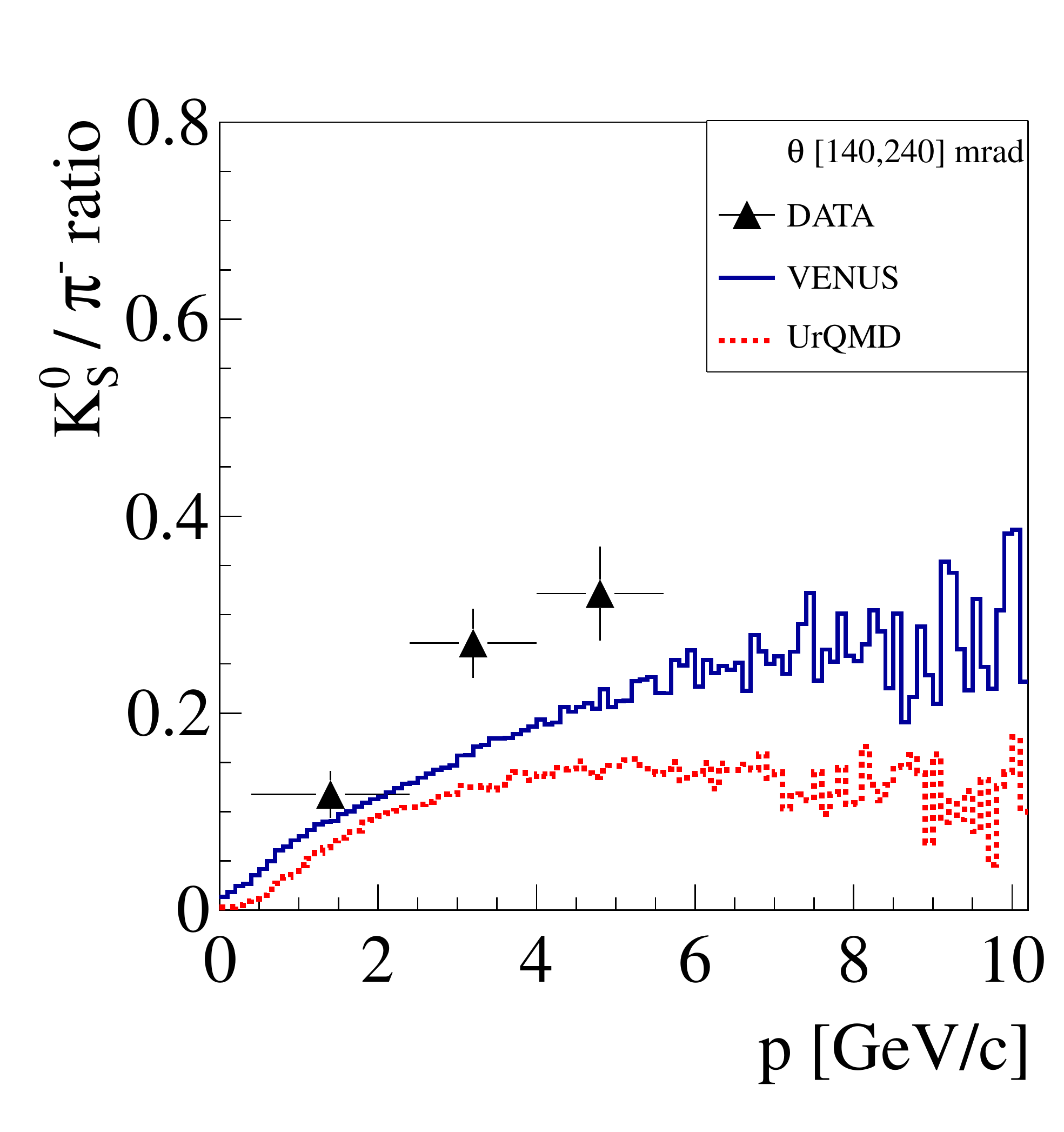}
\end{center}
  \caption{$K^{0}_{S}/\pi^{-}$ ratios versus momentum in two polar angle intervals: [20, 140] mrad (top), [140, 240] mrad (bottom). The vertical error bars on the data points show the total (stat. and syst.) uncertainty. Predictions of two hadron production models are superimposed. }
\label{fig:K0_piminus}
\end{figure}

\begin{figure*}[h!]
\begin{center}
\includegraphics[width=0.45\linewidth]{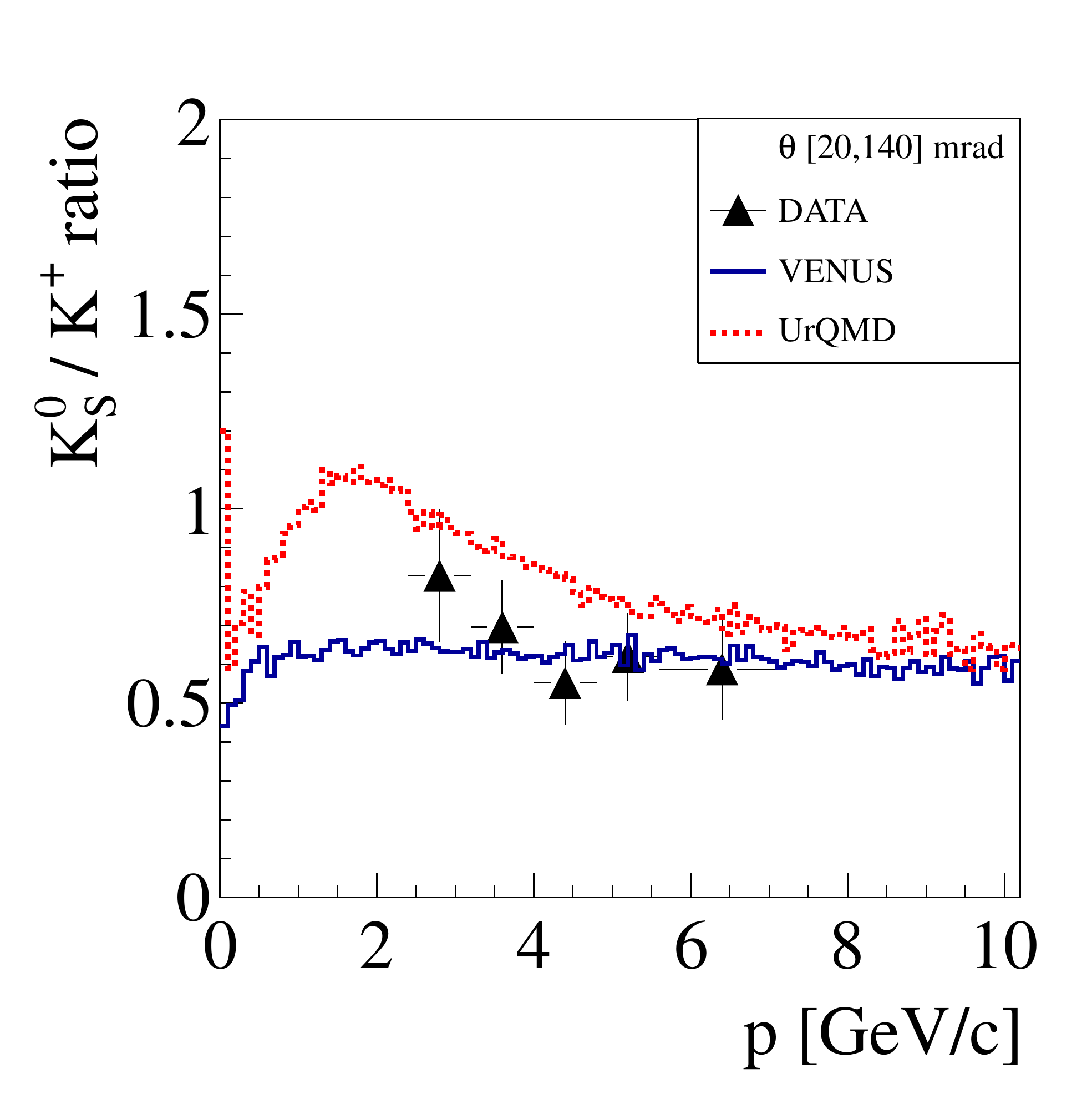}
\includegraphics[width=0.45\linewidth]{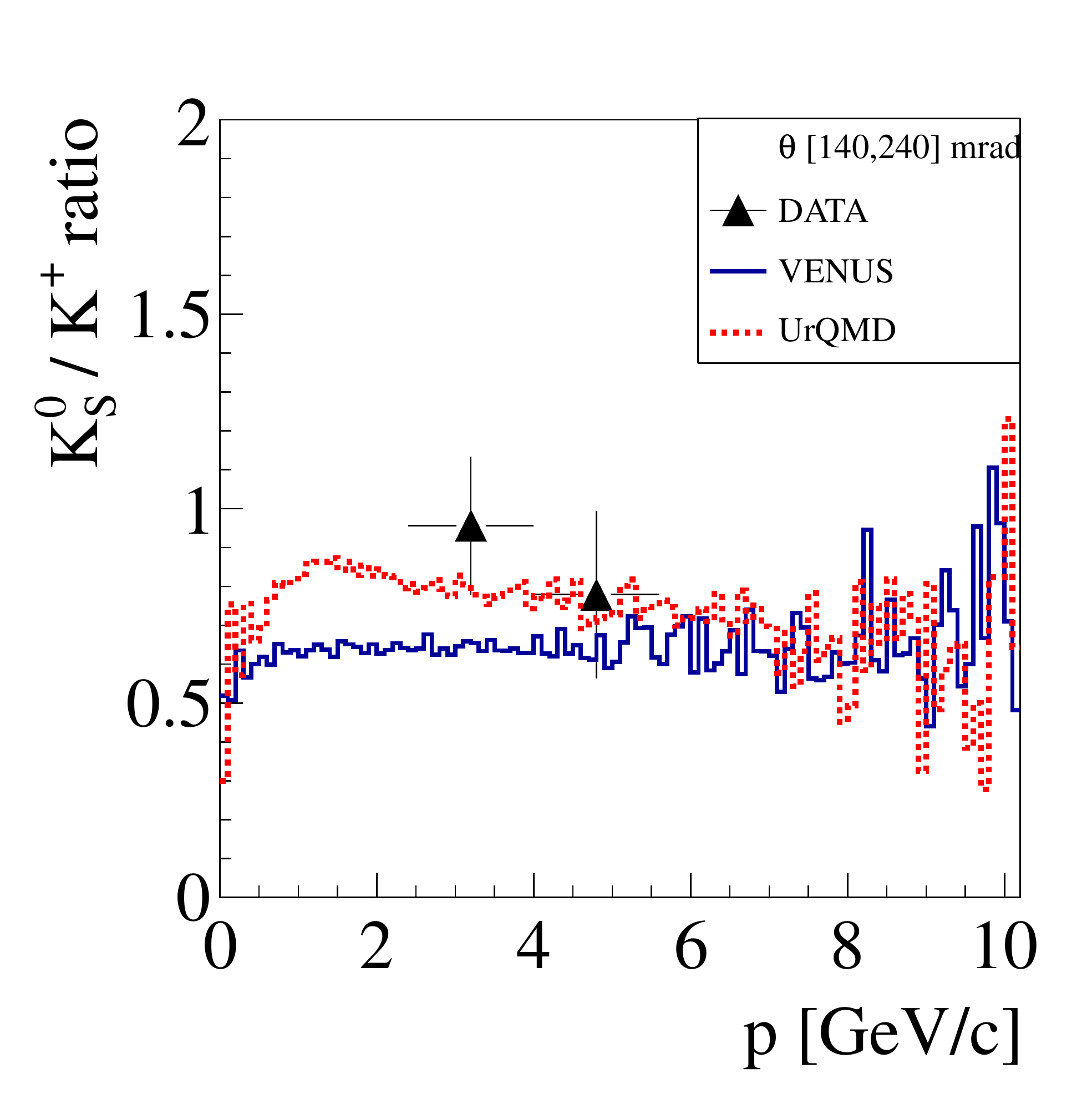}
\end{center}
  \caption{$K^{0}_{S}/K^{+}$ ratios in two polar angle intervals in two polar angle bins 
[20, 140] mrad (top) and [140, 240] mrad (bottom). The vertical error bars on the data points show the total (stat. and syst.) uncertainty. Predictions of hadron production models are superimposed.}
\label{fig:K0_Kplus}
%\end{figure}
\end{figure*}

The $K^{0}_{S}/\pi^{-}$ ratio is plotted versus momentum for selected polar angle intervals in Fig.~\ref{fig:K0_piminus}. 
%\end{figure*}
The $\pi^{-}$ spectra were taken from Ref.~\cite{NA61_pion}.
The ratio of $K^{0}_{S}/\pi^{-}$  is expected to be close to the ratio of the 
$\bar{s}$ to $\bar{u}$ quark production probabilities and it is therefore sensitive to the strangeness  
suppression factor, $\lambda_S$. Furthermore, the ratio $K^{0}_{S}/\pi^{-}$ is used in the calculation of the feed-down corrections when extracting primary pion multiplicities from the data. In Ref.~\cite{NA61_pion} VENUS model predictions, not yet confronted with the NA61/SHINE measurements, were used and therefore large systematic errors were assigned. It is seen from Fig.~\ref{fig:K0_piminus} that the agreement between data and the model is rather satisfactory for the emission angle interval 20-140 mrad whereas the disagreement is large at larger angles. The measured average multiplicity of $K^{0}_{S}$ differs by about 20\% from the one predicted by VENUS and used in the NA61/SHINE correction procedure.
The $K^{0}_{S}/K^{+}$ ratio in the selected polar angle intervals is shown in Fig.~\ref{fig:K0_Kplus}. The $K^{+}$ spectra were measured in the same experiment~\cite{NA61_Kplus} therefore some of the systematic errors canceled out.
%\begin{figure}[h!]
The $d/u$ quark ratio in p+C interactions is 5:7. The smaller yield of $K^{0}_{S}$ compared to that of $K^{+}$ is thus expected because 
it is more probable to find a $u$ current quark to form 
a positively charged kaon ($u\bar{s}$) than a $d$ quark to form a $K^0$ ($d\bar{s}$). 
The $\bar{K}$ ($\bar{d}$s) requires the production of two sea quarks
and thus it is less frequent.

\begin{figure*}[h!]
%\begin{figure}[ht!]
\begin{center}
\includegraphics[width=0.45\linewidth]{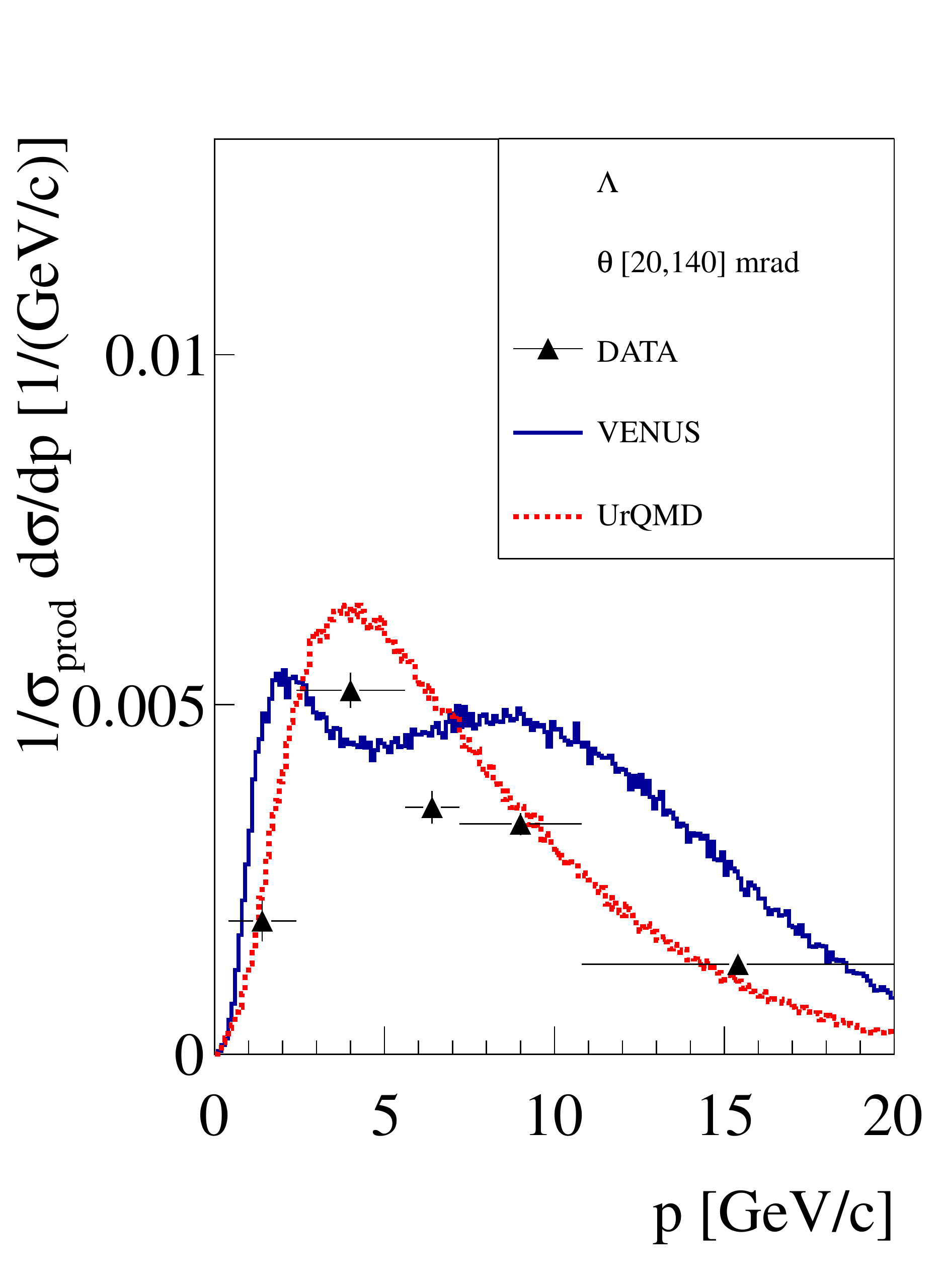}
\includegraphics[width=0.45\linewidth]{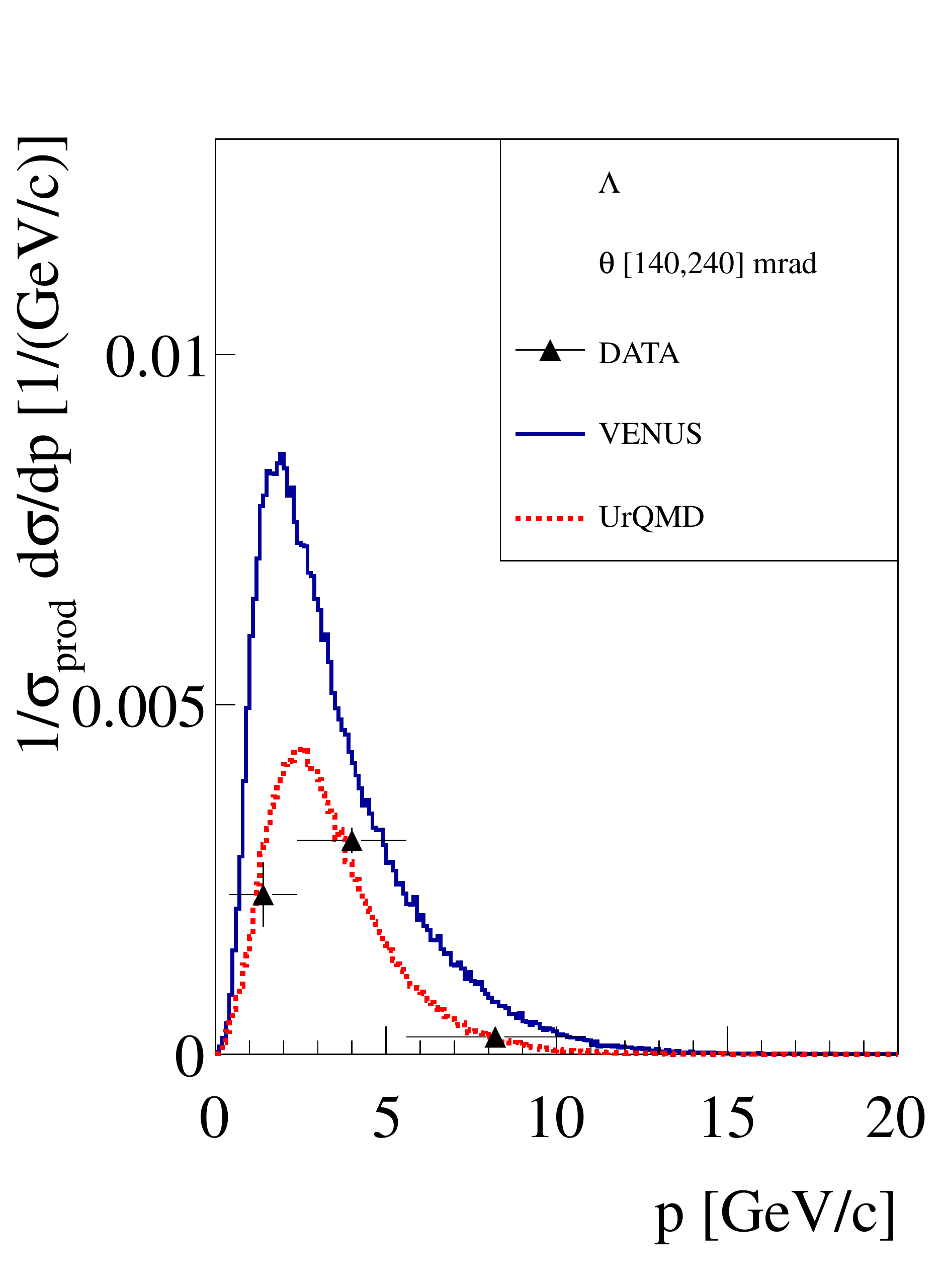}
\end{center}
  \caption{
Mean  multiplicity of $\Lambda$ hyperons in production processes in polar angle intervals: [20, 140] mrad (top), [140, 240] mrad (bottom). Hadron production model predictions are superimposed. Only statistical errors are shown.}
\label{fig:Lambda_dn_dp}
\end{figure*}

The $\Lambda$ results in momentum and polar angle variables normalized 
to mean particle multiplicity in production processes are shown in 
Fig.~\ref{fig:Lambda_dn_dp} with the  predictions from the hadron production models superimposed. 
None of the models provides a satisfactory description of our measurements.
%\end{figure}

%____________________________________________________________________________ 

\clearpage
\section{Summary}
\label{Sec:Summary}

We present the first measurement of the total cross section for $K^{0}_{S}$ production in p+C 
interactions at 31~GeV/c in the experimentally sparsely covered 
region of a few tens of GeV/c beam momentum. 
The increase of the mean $K^0_S$ multiplicity with respect 
to p+p collisions can be explained within errors by a factor 
equal to the average number of primary proton interactions in the carbon nucleus. 
Differential cross sections for $K^{0}_{S}$ and $\Lambda$  production were 
obtained in bins of laboratory momentum and emission angle 
and are compared with predictions of several hadron production models. 
Significant discrepancies between our  
experimental results and predictions of the VENUS 4.12 and UrQMD models are observed. These new 
measurements will help to further refine the prediction of the neutrino beam 
flux in the T2K experiment. 

%____________________________________________________________________________ 

%\clearpage
\section{Acknowledgments}

This work was supported by the Hungarian Scientific Research Fund (grants OTKA 68506 and 71989), 
the Polish Ministry of Science and Higher Education (grants 667/N-CERN/2010/0,  
NN 202 48 4339 and NN 202 23 1837), the National Science Centre of Poland, grant UMO-2012/04/M/ST2/00816,
the Foundation for Polish Science - MPD program, co-financed by the
European Union within the European Regional Development Fund,
the Federal Agency of Education of the Ministry of Education 
and Science of the Russian Federation (grant RNP 2.2.2.2.1547), 
the Russian Academy of Science and the Russian Foundation 
for Basic Research (grants 08-02-00018, 09-02-00664 12-02-91503-CERN), 
the Ministry of Education, Culture, Sports, Science and Technology, Japan, 
Grant-in-Aid for Scientific Research (grants 18071005, 19034011, 19740162, 20740160 and 20039012), 
the German Research Foundation (grants GA 1480/2-1 and GA 1480/2-2), 
the Bulgarian National Scientific Foundation (grant DDVU 02/19/ 2010), 
the Ministry of Education and Science of the Republic of Serbia (grant OI171002), 
the Swiss National fonds Foundation (grant 200020-117913/1) and ETH Research Grant TH-01 07-3. 

Finally, it is a pleasure to thank the European Organization 
for Nuclear Research for a strong support and hospitality and, 
in particular, the operating crews of the CERN SPS accelerator 
and beam lines who made the measurements possible.  

%____________________________________________________________________________ 

%\clearpage
%\addcontentsline{toc}{chapter}{Bibliography}

\include{biblio}

\end{document}